\documentclass[twocolumn,usenatbib]{mnras2}
\pdfoutput=1 %for arXiv submission
\usepackage{amsmath,amstext}
\usepackage[T1]{fontenc}
\usepackage{amssymb}
\input{epsf}
\usepackage{graphicx}
\usepackage{ae,aecompl}

\usepackage{hyperref}
\usepackage{amsmath}
\usepackage{amssymb}
\usepackage{mathtools}
\usepackage{bm}
\usepackage{cleveref}
\usepackage{tensor}
\usepackage{braket}
\usepackage{mhchem}
\usepackage{physics}
\usepackage{upgreek}

\usepackage{color}

\newcommand{\vect}[1]{\mathbf{{#1}}}

\newcommand{\spc}{\quad \quad \quad}

\newcommand{\n}{{\mathrm{n}}}

\def\be{\begin{equation}}
\def\ee{\end{equation}}
\def\beq{\begin{eqnarray}}
\def\eeq{\end{eqnarray}}

\title[Multifluid radiation hydrodynamics]{Multifluid Modelling of Relativistic Radiation Hydrodynamics}
\author[L.~Gavassino et al.]{L.~Gavassino, M.~Antonelli, B.~Haskell\\
Nicolaus Copernicus Astronomical Center, Polish Academy of Sciences, ul. Bartycka 18, 00-716 Warsaw, Poland}
\begin{document}
\maketitle

\begin{abstract}
The formulation of a universal theory for bulk viscosity and heat conduction represents a theoretical challenge for our understanding of relativistic fluid dynamics. Recently, it was shown that the multifluid variational approach championed by Carter and collaborators has the potential to be a general and natural framework to derive (hyperbolic) hydrodynamic equations for relativistic dissipative systems. Furthermore, it also allows keeping direct contact with non-equilibrium thermodynamics, providing a clear microscopic interpretation of the elements of the theory. 
To~provide an example of its universal applicability, in this paper we derive the fundamental equations of the radiation hydrodynamics directly in the context of Carter's multifluid theory. This~operation unveils a novel set of thermodynamic constraints that must be respected by any microscopic model. Then, we prove that the radiation hydrodynamics becomes a multifluid model for bulk viscosity or heat conduction in some  appropriate physical limits.
\end{abstract}


\section{Introduction} 

The hydrodynamic modelling of dissipative systems should guarantee the stability of the homogeneous perfect-fluid states under perturbations. Furthermore, a realistic model for dissipation should be presented in a form which enables an unambiguous contact with microphysics: this~requirement, although not strictly necessary from the mathematical point of view, allows for a clear implementation of microscopic inputs into the macroscopic description. 
For the case of relativistic fluids the problem of finding a theory which fulfils both the requirements is still considered unsolved~\citep{andersson2007review}.
%mdpi: is the bold necessary?
%authors: NO, the bold was intended for the referee (draft version). We remove all the bold.

The natural-looking relativistic generalization of the Navier--Stokes equations~\citep{Weinberg1971}, which~maintains direct contact with the common notions of viscosity and heat conduction, was shown to admit runaway solutions when the homogeneous perfect-fluid states are perturbed~\citep{Hiscock_Insatibility_first_order,gavssino2020arXiv}.  This was shown to be a consequence of the fact that its equations do not admit a well posed initial-value 
problem~\citep{Kost2000}.  Hence, this hydrodynamic description of relativistic viscous fluids is not suitable for computational applications. 
On the other hand, the~second-order theory of viscous fluids proposed by~\cite{Israel_Stewart_1979} introduces some phenomenological coefficients which have a clear microscopic interpretation only in the ideal relativistic gas limit. 
Moreover, there are cases in which the model of Israel and Stewart underestimates the number of non-equilibrium degrees of freedom, so~that it cannot be considered a universal approach to model relativistic dissipative fluids~\citep{BulkGavassino}.

The multifluid formalism of \cite{Carter_starting_point} may be the solution to this long-standing  problem. Partially arising from an action principle, Carter's variational approach leads in a natural way to a well-posed initial value problem governed by hyperbolic equations~\citep{andersson2007review}. 
Thus, its mathematical structure has all the required properties to give rise to a causal and stable theory if the appropriate equation of state is assumed~\citep{Olson1990}. 

Since the multifluid concept was explicitly developed to describe relativistic conducting media~\citep{noto_rel}, Carter's formalism was successfully used as the natural scheme for modelling superfluidity in a covariant framework~\citep{cool1995,Carter_defects2000,gusakovPhysRevD}. 
In this context, the~correspondence with microphysics was completely established~\citep{lebedev1982,carter92,popov2006,Termo}. 
The formalism was applied to the study of the structure of superfluid neutron stars ~\citep{andersson_comer2000,Prix2005,sourie2016PhRvD} and to the formulation of a relativistic theory of vortex dynamics~\citep{langlois98,Prix_single_vortex,Carter_Prix_Magnus} and represents a fundamental tool in relativistic modelling of pulsar glitches~\citep{sourie_glitch2017,antonelli+2018,Geo2020}. 
In addition, it was recently proposed that a multifluid approach might find interesting application in cosmological models~\citep{Osano18,Osano20}.

For the case of relativistic dissipation, a general multifluid model was proposed by \cite{carter1991}. This~model satisfies the conditions of stability and causality (that coincide with the ones of the Israel and Stewart formulation) for small deviations from equilibrium~\citep{Priou1991}. 
Therefore, apart from superfluidity, relativistic multifluids were also studied in the context of heat conduction~\citep{AnderssonLopezFirstOrder,lopez2011,Carter2012}. 
These models, however, still lack of a clear connection with microphysics and are thus not fundamentally preferable to the one of Israel and Stewart.

Recently,  by using arguments of non-equilibrium thermodynamics, it was shown that any bulk-viscous fluid can always be described as a Carter's multifluid if an appropriate choice of thermodynamic variables is adopted~\citep{BulkGavassino}. 
Therefore, at least for the case of dissipation due to bulk viscosity, this result represents a formal justification of the universality of the multifluid formalism and provides a technique for connecting the hydrodynamic model with thermodynamics and kinetic~theory.

In this paper our aim is to provide further insight into the connection of the multifluid theory with microphysics and its universal applicability:
we already discussed the link with the equilibrium thermodynamics of a superfluid~\citep{Termo} and the link with kinetic theory for non-conducting bulk-viscous fluids~\citep{BulkGavassino}. Here, we add another piece to the global picture by studying  how to model a perfect fluid interacting with a radiation fluid within the  Carter multifluid framework. 
Due to its simplicity and wide applicability in astrophysical contexts, this system was widely studied in the literature~\citep{mihalas_book} and the understanding of its properties can be considered satisfactory at every level: statistical mechanics~\citep{huang_book}, kinetic theory~\citep{clayton_book}, thermodynamics~\citep{Leff2002} and hydrodynamics~\citep{rezzolla_book}. 
Therefore, Carter's theory (which may provide a universal hydrodynamic framework) should be able to capture the essential physics of this system. In this sense, investigating the properties of a fluid interacting with radiation in this multifluid framework represents a fundamental test for its descriptive power. 

In addition, it is well known that radiation hydrodynamics admits a diffusion-type limit which reduces the theory to a conventional model for heat conduction~\citep{Shapiro1989,Farris2008}. 
This~implies that the matter-radiation fluid may be the first realistic heat-conducting fluid to be rigorously described in Carter's framework, giving us precious insights about the microscopic origin of the so-called entrainment coupling (a non-dissipative coupling between the species in a multifluid) and the correct implementation of the dissipation coefficients. 

Throughout the paper we adopt the spacetime signature $ ( - , +, + , + ) $ and work in natural units $c=G=k_B=1$. Moreover, the~symbol $\upgamma$ is used to label the quantities related to the photon fluid and should not be interpreted as a spacetime index. 

\section{Multifluid Hydrodynamics}

We briefly review the basic ideas of the multifluid formalism. The~general theory was formulated in~\cite{noto_rel} and~\cite{Carter_starting_point}, see also~\cite{andersson2007review} for a review. 
The formalism was also extended to incorporate shear viscosity~\citep{carter1991} and elasticity~\citep{ander_elasticity_2019} but these effects will not be considered in the present work.

\subsection{Non-Dissipative Evolution of Relativistic Multifluids}\label{polj}  

The variational approach of Carter builds on the assumption that one can identify a set of four-currents $n_x^\nu$ describing different flows in the system. 
Given the scalars
\begin{equation}
n_{xy}^2 := -n_x^\nu n_{y\nu},
\end{equation}
an equation of state for the fluid must be provided in terms of a Lagrangian density
\begin{equation}\label{gggg}
\Lambda = \Lambda(n_{xy}^2)  \spc  x \leq y.
\end{equation}

The condition $x \leq y$ is imposed to avoid repeated arguments, since $n_{xy}^2 = n_{yx}^2$.
Introducing the bulk coefficients
\begin{equation}\label{Bb}
\mathcal{B}^{x} := -2 \dfrac{\partial \Lambda}{\partial n_{xx}^2}  
\end{equation}
and the anomalous coefficients
\begin{equation}\label{Aa}
\mathcal{A}^{xy}:= - \dfrac{\partial \Lambda}{\partial n_{xy}^2} =:\mathcal{A}^{yx}  \spc x < y,
\end{equation} 
we can define the conjugate momenta of the currents as 
\begin{equation}\label{vbvbvb}
\mu^x_\nu := \dfrac{\partial \Lambda}{\partial n_x^\nu} = \mathcal{B}^x n_{x\nu} + \sum_{y \neq x} \mathcal{A}^{xy} n_{y\nu}.
\end{equation}

The anomalous coefficients (whose presence in a multifluid is the norm) incorporate the entrainment effect, a non-dissipative coupling between the currents. Historically, the~importance of entrainment was first recognized in the context of superfluid mixtures~\citep{khala57,andreevbashkin1976}, but it is a general feature of Carter's variational approach.

In the literature it is common to find an alternative procedure of differentiating the Lagrangian density which includes also the terms with $x>y$ and treats $n_{xy}^2$ and $n_{yx}^2$ as independent variables, see~e.g.,~\cite{Carter_starting_point,Prix_single_vortex}. 
We discuss the connection with the present approach in Appendix \ref{AAA}.

A non-dissipative hydrodynamic model can be obtained by considering an action of the form
\begin{equation}\label{action}
I = \int \bigg( \dfrac{R}{16 \pi} +\Lambda \bigg) \sqrt{-g} \, d^4 x,
\end{equation}
where $R$ is the scalar curvature and $\sqrt{-g}$ is the square root of the absolute value of the determinant of the metric. The~domain of the action is set imposing that the currents are conserved,
\begin{equation}\label{consSSSssS}
\nabla_\nu n_x^\nu =0,
\end{equation}
both on-shell and off-shell. To make sure that this is indeed satisfied, the~variations of the currents are taken in the Taub form~\citep{taub_1954}
\begin{equation}
\delta n_x^\nu = \xi_x^\rho \nabla_\rho n_x^\nu -n_x^\rho \nabla_\rho \xi_x^\nu + n_x^\nu \bigg( \nabla_\rho \xi_x^\rho - \dfrac{1}{2} g^{\rho \sigma} \delta g_{\rho \sigma} \bigg),
\end{equation} 
where the vector field $\xi_x^\nu$ describes an arbitrary infinitesimal displacement of the world-lines of the fluid elements of the species $x$. 
The Euler-Lagrange equations are obtained imposing the stationarity of the action with respect to arbitrary infinitesimal variations $\delta g_{\nu \rho}$  and displacements $\xi_x^\nu$. 
The first one produces Einstein's equations,
\begin{equation}\label{Einstein}
G^{\nu \rho} = 8 \pi T^{\nu \rho},
\end{equation}
where the energy momentum tensor has the form
\begin{equation}\label{Tnuro}
T\indices{^\nu _\rho} = \Psi \delta\indices{^\nu _\rho} + \sum_x n_x^\nu \mu^x_\rho.
\end{equation}

The scalar $\Psi$ can be interpreted as a generalised thermodynamic pressure and is given by
\begin{equation}
\label{preSSSSSSSSSSSSSSSSSSSSSUUUUUUUUUUre}
\Psi = \Lambda - \sum_x n_x^\nu \mu_\nu^x.
\end{equation}

Ignoring the boundary terms which do not contribute to the equations of motion,
the variation of the action produced by the displacements $\xi_x^\nu$ has the form
\begin{equation}\label{hyihyihyi}
\delta I = \int \bigg( \sum_x f^x_\nu \xi_x^\nu \bigg) \sqrt{-g} \, d^4 x,
\end{equation}
where 
\begin{equation}
f^x_\nu := 2n_x^\rho \nabla_{[ \rho} \mu^x_{\nu]}
\end{equation}
can be interpreted as the force per unit volume acting to the species $x$. The~ condition $\delta I=0$ for any independent choice of $\xi_x^\nu$ produces the Euler-Lagrange equations
\begin{equation}\label{forces}
f_\nu^x =0 \spc \forall x.
\end{equation} 

Equations \eqref{consSSSssS}, \eqref{Einstein} and \eqref{forces} constitute a system which arises from a well posed action principle  and, therefore, are given in the form of an initial value problem~\citep{andersson2007review}. 
Note that from \eqref{consSSSssS} and \eqref{Tnuro}, one~can show that
\begin{equation}
\nabla_\rho T\indices{^\rho _\nu} = \sum_x f^x_\nu.
\end{equation}

However, taking the divergence of \eqref{Einstein}, one immediately has the energy-momentum conservation
\begin{equation}\label{conservation}
\nabla_\rho T\indices{^\rho _\nu} =0.
\end{equation}

Therefore even in the case in which the forces $f^x_\nu$ were not zero, their sum must vanish,
\begin{equation}
\sum_x f^x_\nu =0,
\end{equation}
which is Newton's third law. The~formalism is easily extended to the case in which there are some currents that are locked to each other. For example,  assume that the species $x$ and the species $y$ interact, and are coupled on time-scales much shorter than those we are interested in. In this case, we can take them to be at rest with respect to each other. The~motion is still adiabatic, but now it is subject to the geometrical constraint
\begin{equation}\label{rgrg}
n_x^{[\nu} n_y^{\rho]} =0.
\end{equation}

This~is implemented by imposing the world-line displacements $\xi_x^\nu$ and $\xi_y^\nu$ to satisfy the constraint
\begin{equation}\label{sxsxsx}
\xi_x^\nu = \xi_y^\nu.
\end{equation}

Thus, from the variation \eqref{hyihyihyi} we find that $f^x_\nu$ and $f^y_\nu$ do not need to vanish separately, but the total force density does,
\begin{equation}\label{kkkkkil}
f^x_\nu + f^y_\nu =0.
\end{equation}
Again, this condition is Newton's third law. 

\subsection{Including Dissipation}\label{InclDissip12344567889}

The second law of thermodynamics is not automatically provided by the action principle, so~that  the dissipative terms of the theory have to be supplied in some other way and inserted by hand into the equations of motion. However, the~study of the adiabatic regimes presented in the previous subsection can be used as a guideline for a consistent inclusion of these additional terms.

The current conservation \eqref{consSSSssS} in a dissipative regime may not hold, but chemical-type transfusions may be allowed. Therefore, we need to replace Equation~\eqref{consSSSssS} with
\begin{equation}\label{ratesqwe}
\nabla_\nu n_x^\nu = r_x,
\end{equation}
where $r_x$ describes the rate (per unit volume and time) of production of the species $x$. 
The second law of thermodynamics is implemented by considering an additional current $s^\nu$ 
(interpreted as the entropy current) whose production rate must satisfy the constraint
\begin{equation}
r_s = \nabla_\nu s^\nu \geq 0 \, .
\end{equation}

The energy-momentum tensor is assumed to maintain the form \eqref{Tnuro} and we require it to still satisfy Einstein's equations. Now, its four-divergence takes the form
\begin{equation}
\nabla_\rho T\indices{^\rho _\nu} = \sum_x \mathcal{R}^x_\nu,
\end{equation}
where
\begin{equation}\label{Rx}
\mathcal{R}^x_\nu = f^x_\nu + r_x \mu^x_\nu
\end{equation}
represent the dissipative generalization of the Lagrangian forces $f^x_\nu$. 
Comparison with \eqref{Einstein} tells that Newton's third law is still valid,
\begin{equation}\label{3newt}
\sum_x \mathcal{R}^x_\nu =0,
\end{equation}
but the terms $\mathcal{R}^x_\nu$ do not need to vanish separately. 

The quantities $r_x$ and $\mathcal{R}^x_\nu$ incorporate dissipation in the theory and have to be modelled according to microphysical arguments. Now, consider Equation~\eqref{Rx} for $x=s$,
\begin{equation}
\mathcal{R}^s_\nu = f^s_\nu + r_s \Theta_\nu,
\end{equation}
where we adopted the notation  $\Theta_\nu := \mu_\nu^s$ for the conjugate momentum to the entropy current. 
Contracting with the four-velocity
\begin{equation}
u_s^\nu := \dfrac{s^\nu}{\sqrt{-s_\rho s^\rho}},
\end{equation}
and using Equation~\eqref{3newt} we find
\begin{equation}\label{entrprod}
 r_s =  \dfrac{1}{\Theta_s}  \sum_{x \neq s} u_s^\nu \mathcal{R}^x_\nu \geq 0,  
\end{equation}
where we introduced the quantity
\begin{equation}\label{ThetaS}
\Theta_s := -\Theta_\nu u_s^\nu.
\end{equation}

Thus, only the coefficients $r_x$ and $\mathcal{R}^x_\nu$ for $x \neq s$ need to be computed from microphysical calculations. Then, $r_s$ and $\mathcal{R}^s_\nu$ are  obtained through the identities \eqref{3newt} and \eqref{entrprod}.

Finally, there is a subtlety we need to remark on. We have introduced the forces $\mathcal{R}^x_\nu$ as dissipative contributions, but from \eqref{entrprod} we see that this is not strictly necessary. 
In fact, one may in principle design them in such a way that
\begin{equation}\label{vanishDissipPa}
\sum_{x \neq s} u_s^\nu \mathcal{R}^x_\nu =0
\end{equation}
is guaranteed by construction, without requiring that the forces themselves vanish. 
Under this condition, no entropy can be produced and the theory is still non-dissipative. 
We do not consider this possibility in the following. However, we will be forced to come back to discuss this point in greater detail  in Section~\ref{theHYDRA}.

\section{Heat Conduction and Bulk Viscosity}\label{HTBV}

Relativistic models for heat conduction and bulk viscosity naturally arise as particular cases of the general multifluid theory. In this section, we briefly summarize some results that were obtained up to now. 

\subsection{Heat Conduction} \label{HCX1}

Consider a fluid comprised of indistinguishable particles of a single type, whose number four-current $n^\nu$ is conserved,
\begin{equation}\label{kikikik}
\nabla_\nu n^\nu =0.
\end{equation}

In the presence of heat conduction, the~entropy current $s^\nu$ is generally not aligned with the particle flux.  Therefore, we consider a minimal two-fluid model with two independent currents, $n^\nu$ and $s^\nu$. The~Lagrangian density takes the form
\begin{equation}\label{zs}
\Lambda = \Lambda (n^2,n_{ns}^2,s^2),
\end{equation}
whose differential is
\begin{equation}
d\Lambda = -\dfrac{\mathcal{B}}{2} d(n^2) - \mathcal{A} d(n_{ns}^2) - \dfrac{\mathcal{C}}{2} d(s^2).
\end{equation}

The conjugate momenta $\mu_\nu$ and $\Theta_\nu$, to particle and entropy current respectively, are
\begin{equation}
\begin{split}\label{euqation34}
& \mu_\nu = \mathcal{B}n_\nu + \mathcal{A}s_\nu 
\\
& \Theta_\nu = \mathcal{C}s_\nu + \mathcal{A}n_\nu .  
\end{split}
\end{equation}

The pressure reads
\begin{equation}\label{Press}
\Psi = \Lambda - n^\nu \mu_\nu - s^\nu \Theta_\nu
\end{equation}
and the energy momentum tensor takes the form
\begin{equation}\label{decomp}
T\indices{^\nu _\rho} = \Psi \delta \indices{^\nu _\rho} + n^\nu \mu_\rho + s^\nu \Theta_\rho .
\end{equation}

This~energy-momentum tensor might not look similar to the ones adopted in conventional models for heat conduction, but it shares the same geometrical structure. 
This~is more easily seen by working in the Eckart frame, namely by introducing the fluid four-velocity as
\begin{equation}
u_n^\nu = \dfrac{n^\nu}{\sqrt{-n_\rho n^\rho}}
\end{equation}
and defining the quantities
\begin{equation}\label{Ecktemp}
s_E := -u_{n\nu} s^\nu  
\spc 
\Theta_E := - \Theta_\nu u_n^\nu,
\end{equation} 
where the first is the entropy density measured in the frame defined by the particle current. 
Now,~the~heat-flux $q^\nu$ is given by means of the orthogonal decomposition
\begin{equation}\label{entrcurr}
s^\nu = s_E u_n^\nu + \dfrac{q^\nu}{\Theta_E}  \spc  q_\nu u_n^\nu =0.
\end{equation}

By defining the internal energy density as the one measured in the frame of the particle current, 
\begin{equation}
\mathcal{U}= T_{\nu \rho} u_n^\nu u_n^\rho,
\end{equation}
and the coefficient
\begin{equation}\label{DDD}
\mathcal{D} =\dfrac{\mathcal{C}}{\Theta_E^2},
\end{equation}
it is possible to show that the expression \eqref{decomp} decomposes into
\begin{equation}\label{Eckart}
T^{\nu \rho} = 
(\mathcal{U}+\Psi)u_n^\nu u_n^\rho + \Psi g^{\nu \rho} 
+ 2 q^{(\nu} u_n^{\rho)} + \mathcal{D}q^\nu q^\rho .
\end{equation} 

At the first order in the heat flux, the~above formula reduces to the energy-momentum tensor of the Eckart heat-conducting fluid~\citep{Hiscock_Insatibility_first_order}. 
The additional term $\mathcal{D}q^\nu q^\rho$ is associated with the fact that the flux of energy introduces an anisotropy along its direction which might in principle have an effect on the stress tensor. However, since the stress tensor has to be invariant under the transformation $q^\nu \rightarrow -q^\nu$, this correction is second order in the heat flux. 

The dissipative hydrodynamic equations take the form
\begin{equation}\label{jujujuj}
\begin{split}
& \mathcal{R}^n_\nu = f^n_\nu \\
& \mathcal{R}^s_\nu = f^s_\nu + r_s \Theta_\nu , 
\end{split}
\end{equation}
where we have used Equation~\eqref{kikikik} to set $r_n=0$. The~dissipative tensors which have to be provided by studying the microphysics of the system are $\mathcal{R}^n_\nu$, $\mathcal{R}^s_\nu$ and $r_s$. 
However, they are not all independent; in~fact from \eqref{3newt} and \eqref{entrprod} we have that 
\begin{equation}
 \mathcal{R}^s_\nu =-\mathcal{R}^n_\nu    \spc  r_s = \dfrac{u_s^\nu \mathcal{R}^n_\nu}{\Theta_s},
\end{equation}   
thus we only need to determine $\mathcal{R}^n_\nu$. It is possible to further reduce the number of unknowns by means of geometrical arguments. In the simplest model, proposed by \cite{noto_rel}, it is assumed that $\mathcal{R}^n_\nu$ is a function of the currents $n^\nu$ and $s^\nu$ only (i.e.,~not of their derivatives). 
Then the force assumes the form
\begin{equation}\label{frct}
\mathcal{R}^n_\nu =\alpha q_\nu,
\end{equation}
where
\begin{equation}
\alpha=\alpha(n^2,n_{ns}^2,s^2) \geq 0,
\end{equation}
is a transport coefficient to be determined from kinetic theory.
Equation \eqref{frct} can be derived from the fact that by isotropy, $\mathcal{R}^n_\nu$ is a linear combination of $s^\nu$ and $n^\nu$ and, from the first equation of \eqref{jujujuj}, needs to be orthogonal to $u_n^\nu$. 
The positivity of $\alpha$ is ensured by the equation
\begin{equation}
\label{propriocosi}
r_s =  \dfrac{\alpha q_\nu q^\nu}{s \Theta_s \Theta_E } \geq 0,
\end{equation}
where we have assumed that $\Theta_E$ and $\Theta_s$ are positive. In fact, they reduce to the usual notion of temperature at equilibrium, so this is equivalent to assume that the system is sufficiently close to thermodynamic equilibrium (out of equilibrium a rigorous definition of temperature does not exist and only on equilibrium states $\Theta_E$ and $\Theta_s$ both coincide with the thermodynamic temperature).
%mdpi: Footnote is not accepted in this journal, we removed all of them into main text, please confirm. 
%authors: We modified in order to incorporate the footnotes into the main text.

Equation \eqref{frct} models the force $\mathcal{R}^n_\nu$ as a viscous friction between the particle current and the entropy current. Clearly, the~effect of such a friction is to drive the system towards a state in which the entropy and the particles flow together. 
Alternative models for $\mathcal{R}^n_\nu$ were proposed, which~include terms involving also the derivatives of the hydrodynamic quantities~\citep{lopez2011}. 
As a result, in this case the force also has a component which is orthogonal to both $n^\nu$ and $s^\nu$. For small deviation from equilibrium the two models coincide and both reduce to the one of~\cite{Israel_Stewart_1979}. 
In this section, we adopt the model of~\cite{noto_rel} for its simplicity, but the possible existence of terms which contain derivatives cannot be ruled out in principle. We will come back to this point in Section~\ref{theHYDRA}.

In both cases, this system of equations was shown to have the structure of a relativistic Cattaneo equation~\citep{cattaneo1958,AnderssonLopezFirstOrder,lopez2011}. It is given in a form which is naturally hyperbolic, and therefore compatible with causality, and it becomes a good model for the second sound for high frequency perturbations~\citep{rezzolla_book}. 

When perturbations are slow, i.e.,~evolve on timescales that are longer than the characteristic  relaxation time-scale~\citep{AnderssonLopezFirstOrder}
\begin{equation}
\tau_r =  \dfrac{\mathcal{C}s\Theta_s}{\alpha \Theta_E^2} ,
\end{equation}
the conventional Navier-Stokes model for heat conduction is recovered. In the Navier-Stokes limit of Carter's model the thermal conductivity coefficient is given by 
\begin{equation}
\kappa = \dfrac{s\Theta_s}{\alpha \Theta_E},
\end{equation}
so that Formula \eqref{propriocosi}, for the entropy production, acquires the more familiar form
\begin{equation}\label{kufkuf}
r_s =  \dfrac{ q_\nu q^\nu}{\kappa \, \Theta^2_E } .
\end{equation}

\subsection{Bulk Viscosity}

Bulk viscosity arises from the fact that the fluid has internal degrees of freedom which go out of equilibrium due to expansion and contraction of the volume elements in the hydrodynamic evolution.  
These degrees of freedom can always be modelled as additional currents $n_A^\nu$, $A=1,...,l-1$, which~are locked to the conserved particle current $n^\nu$, provided that the volume element is locally isotropic (which implies the absence of shear viscosity and heat conduction) in the particle rest-frame~\citep{BulkGavassino}.  Hence,~a~bulk-viscous fluid can always be modelled as a multifluid whose currents $n^\nu$, $s^\nu$ and $n_A^\nu$ are all subject to the geometrical constraint \eqref{rgrg}. 
In this case, the Lagrangian density $\Lambda$ reduces to 
\begin{equation}\label{bulkio}
\Lambda = - \mathcal{U},
\end{equation}
where $\mathcal{U}$ is the internal energy of the fluid. Its differential is
\begin{equation}\label{labulkcomelasoio}
d\mathcal{U} = \mu dn + \Theta ds -\mathbb{A}^A dn_A,
\end{equation}
where we use the Einstein summation convention for the chemical index $A=1,...,l-1$. 
It is easy to prove that
\begin{equation}
\mu_\nu = \mu u_\nu  \spc \Theta_\nu = \Theta u_\nu  \spc \mu_\nu^A = -\mathbb{A}^A u_\nu,
\end{equation}
where $u^\nu$ is the (unique) four-velocity of this non-conducting multifluid. Moreover, the~pressure and the energy-momentum tensor take the familiar perfect fluid  forms
\begin{equation}\label{pressandenem}
\begin{split}
& \Psi = -\mathcal{U} +n\mu + s\Theta -\mathbb{A}^A n_A \\
& T\indices{^\nu _\rho} = \Psi \delta\indices{^\nu _\rho} + (\mathcal{U}+\Psi)u^\nu u_\rho  . 
\end{split}
\end{equation}

The symbols $-\mathbb{A}^A$ are adopted for the chemical potentials of the species $A$ because the currents $n_A^\nu$ exist only as a parametrization of the out-of-equilibrium states (see section II-B of~\cite{BulkGavassino}). In local thermodynamic equilibrium, according to the minimum energy principle~\citep{Callen_book}, they have the value which minimizes the energy at fixed $n$ and $s$, and this gives rise to the condition
\begin{equation}\label{formuloio}
\mathbb{A}^A =0.
\end{equation}

For this reason the $\mathbb{A}^A$ can be interpreted as generalised chemical affinities, justifying the adopted~notation.

The collinearity condition \eqref{rgrg}, valid for all the currents, simplifies the equations of motion considerably. The~independent rates which need to be provided by microphysics are the coefficients $r_A$, which near equilibrium, can be expanded to the linear order in the affinities,
\begin{equation}\label{uiop}
r_A = \Xi_{AB}\mathbb{A}^B.
\end{equation}

The $(l-1)\times(l-1)$ matrix $\Xi_{AB}$ is symmetric as a result of Onsager's principle. The~remaining equations of motion, which~are needed to completely specify the hydrodynamic evolution,  are  given by the particle and energy-momentum conservation
\begin{equation}\label{hjkl}
 r_n=0  \spc  \nabla_\rho T\indices{^\rho _\nu}=0 . 
\end{equation}

Equation \eqref{uiop} and \eqref{hjkl} can be combined, giving the formula for the entropy production:
\begin{equation}
\Theta r_s = \mathbb{A}^A r_A = \Xi_{AB} \mathbb{A}^A \mathbb{A}^B \geq 0,
\end{equation}
which implies that $\Xi_{AB}$ is definite non-negative (strictly positive if the ergodic assumption is made, see e.g.,~\cite{khinchin_book}).

The model we have presented is constructed in a form that is naturally hyperbolic, and therefore it is compatible  with the basic requirement necessary for causality and stability. 
It was shown in~\cite{BulkGavassino} that when $l=2$ the above model reduces, for small deviations from equilibrium, to the bulk viscosity prescription derived by Israel and Stewart, which~is known to be (conditionally) causal and stable.

When the hydrodynamic evolution is slow enough compared to the microscopic equilibration timescales, the~model reduces to a relativistic Navier-Stokes description of bulk viscosity  (see~sections~II-D and VII in~\cite{BulkGavassino}), with a bulk viscosity coefficient given by
\begin{equation}\label{ZZZeTa}
\zeta = \Xi^{AB} \dfrac{\partial x_A^{\text{eq}}}{\partial v} \bigg|_{x_s} \dfrac{\partial x_B^{\text{eq}}}{\partial v} \bigg|_{x_s}.
\end{equation} 

Here the matrix $\Xi^{AB}$ is the inverse of $\Xi_{AB}$, while $x_s=s/n$ is the entropy per particle and
\begin{equation}
x_A^{\text{eq}}(v,x_s) = \dfrac{n_A}{n} \bigg|_{\mathbb{A}^B =0} ,
\end{equation}
is the equilibrium fraction of the effective chemical species labelled by $A$.

It is, finally, important to remark a subtlety about the dissipation in a multifluid context. As can be seen from the foregoing discussion, in a multifluid approach, heat conduction and bulk viscosity are not implemented directly as small corrections to the stress-energy tensor but they are modelled in a non-perturbative way by introducing further (non-equilibrium) degrees of freedom in the theory. The~immediate consequence is that heat conduction (i.e.,~the flow of energy in the matter's rest-frame) and bulk viscosity (i.e.,~the non-equilibrium correction to the pressure) are, in a generic multifluid, interconnected (influencing each other at every order, higher than the first~\citep{Israel_Stewart_1979})  and cannot be completely separated. For this reason, in the present paper, we have introduced the purely heat-conducting fluid and the purely bulk-viscous fluid separately, while in principle a generic multifluid will contain both the processes.

\section{Radiation Hydrodynamics}\label{RHfoundation}

We show that the equations of the radiation hydrodynamics in the $M_1$ closure scheme~\citep{Levermore1984,Sadowski2013,Fragile2014} can be conveniently obtained directly from Carter's multifluid formalism. This~alternative derivation provides considerable thermodynamic and geometrical insight. 

Our study will be specifically devoted to photon radiation, so~that we use  the label $\upgamma$ to indicate the quantities related to the radiation fluid. We remark, however, that the following discussion holds in principle also for any kind of radiation which does not carry any conserved charge (for example, it would apply also to the case of  a real scalar boson or a Majorana fermion). In a thermodynamic perspective, this condition corresponds to the requirement that the exchange of radiation between matter elements is a pure phenomenon of heat transfer~\citep{Gavassino2020Zeroth} and not a chemical transfusion.

Although Carter's formalism could be used to model also neutrino radiation (which will be studied in detail in future work), such a system would not admit a chemical-equilibrium limit that is a model of heat conduction. Instead, it would become a charge-conducting fluid, where the transported charge is the lepton number. 
Therefore, the~presence of a conserved charge associated with the radiation fluid is the  reason  neutrino radiation is physically different from the photon case.

\subsection{The Hydrodynamic Model}

We consider systems containing two particle currents, $n^\nu$ and $\upgamma^\nu$ (to avoid confusion, the~labels $n$ and $\upgamma$ will always be used as chemical labels and never as space-time indices).
The first is assumed to be an exactly conserved current,
\begin{equation}\label{frftgthyh}
r_n = \nabla_\nu n^\nu =0,
\end{equation}
and represents the flow of the matter component of the multifluid. The~second, namely $\upgamma^\nu$, is the  current density associated with photons, whose number is not conserved (it can change in absorption and emission processes),
\begin{equation}\label{rgaMMa}
r_\upgamma = \nabla_\nu \upgamma^\nu \neq 0 .
\end{equation}

We impose Boltzmann's  molecular chaos  ansatz~\citep{huang_book}, namely that the statistical correlations between matter and radiation can be neglected. This~allows defining two separate entropy currents $s_n^\nu$ and $s_\upgamma^\nu$ associated with the matter and the radiation, whose sum gives the total entropy current~\citep{Groot1980RelativisticKT}:
\begin{equation}\label{dfgtyh}
s^\nu = s_n^\nu +s_\upgamma^\nu.
\end{equation} 
Note that the second law requires  
\begin{equation}\label{dddddddri}
\nabla_\nu s^\nu=\nabla_\nu s_n^\nu+\nabla_\nu s_\upgamma^\nu \geq 0,
\end{equation}
but the two entropies do not need to grow separately. 

To simplify the system we impose that the heat conduction parameters of matter and radiation vanish, namely
\begin{equation}\label{uytgbnm}
n^{[\nu}s_n^{\rho]}=0  \spc \upgamma^{[\nu}s_\upgamma^{\rho]}=0 .
\end{equation} 

Finally, we assume that the interactions between matter and the radiation have the form of local collision processes, which~occur for sufficiently short times that the statistical average of the interaction term of the microscopic Hamiltonian can be neglected. 
This~allows us to decompose the Lagrangian density into a matter and a radiation part, for which we will adopt a simple separability prescription 
\begin{equation}
\label{LLambDDaaAqwerty}
\Lambda = -\rho(n,s_n) - \varepsilon(\upgamma,s_\upgamma),
\end{equation}
where $\rho$ is a pure function of $n=\sqrt{-n_\nu n^\nu}$ and $s_n=\sqrt{-s_{n\nu}s_n^{\nu}}$, while $\varepsilon$ is a pure function of $\upgamma=\sqrt{-\upgamma_\nu \upgamma^\nu}$ and $s_\upgamma=\sqrt{-s_{\upgamma \nu}s^\nu_{\upgamma}}$.

By comparison with \eqref{bulkio} we interpret $\rho$ as the internal energy of the matter fluid measured in its own rest-frame, that is identified by the four-velocity $u_n^\nu = n^\nu/n$. 
Therefore, its differential takes the~form
\begin{equation}\label{dRho}
d\rho = \Theta^n ds_n + \mu dn,
\end{equation}
where $\Theta^n$ and $\mu$ are the temperature and the chemical potential of the fluid. Analogously,  $\varepsilon$ is the internal energy of the radiation fluid, measured in the frame defined by $u_\upgamma^\nu = \upgamma^\nu/\upgamma$, and its differential has the form 
\begin{equation}\label{dVarepsilon}
d\varepsilon = \Theta^\upgamma ds_\upgamma -\mathbb{A}^\upgamma d\upgamma,
\end{equation}
where $\Theta^\upgamma$ is the temperature of the radiation fluid. 
In the above equation we introduced the notation $\mu^\upgamma =-\mathbb{A}^\upgamma$ to recall that in thermodynamic equilibrium the chemical potential of the radiation fluid $\mu^\upgamma$ must vanish. 
In particular, if the emission/absorption process 
\begin{equation}
n \ce{ <=> } n+\upgamma,
\label{cucciolottobello}
\end{equation} 
is interpreted as a chemical-type reaction between the matter and radiation, the~ affinity $\mathbb{A}^\upgamma$ associated with the above reaction is minus the chemical potential of photons. 

Note that the  reaction \eqref{cucciolottobello} is possible only because the radiation does not carry any conserved charge. For neutrino radiation this is no longer the case, due to the conservation of the lepton number. As a result, the~neutrino chemical potential does not vanish in chemical equilibrium~\citep{Mazurek1975} and the present discussion does not apply.

It is possible to show that the conjugate momenta to the currents $s_n^\nu$, $n^\nu$, $s_\upgamma^\nu$ and $\upgamma^\nu$ are, respectively,
\begin{equation}\label{moMenti}
\begin{split}
& \Theta_\nu^n = \Theta^n u_{n\nu} \spc \mu_\nu = \mu u_{n\nu} \\
& \Theta_\nu^\upgamma = \Theta^\upgamma u_{\upgamma\nu} \spc -\mathbb{A}^\upgamma_\nu = -\mathbb{A}^\upgamma u_{\upgamma\nu}.
\end{split} 
\end{equation} 

The generalised thermodynamic pressure given in \eqref{preSSSSSSSSSSSSSSSSSSSSSUUUUUUUUUUre} splits into 
\begin{equation}\label{kopf}
\Psi = P_n + P_\upgamma ,
\end{equation}
where
\begin{equation}
P_n = -\rho + \Theta^n s_n + \mu n
\end{equation}
is the pressure of the matter fluid, while
\begin{equation}\label{PRRraDD}
P_\upgamma = -\varepsilon + \Theta^\upgamma s_\upgamma -  \mathbb{A}^\upgamma \upgamma
\end{equation}
is the pressure of the radiation fluid. Thus, the~variational principle presented in Section~\ref{polj} leads to a completely decoupled energy-momentum tensor
\begin{equation}\label{TTT}
T^{\nu \rho} = M^{\nu \rho} + R^{\nu \rho},
\end{equation}
where
\begin{equation}\label{perfettiloro}
\begin{split}
& M^{\nu \rho} = (\rho +P_n)u_n^\nu u_n^\rho + P_n g^{\nu \rho} \\
& R^{\nu \rho} = (\varepsilon +P_\upgamma)u_\upgamma^\nu u_\upgamma^\rho + P_\upgamma g^{\nu \rho} 
\end{split}
\end{equation}
are respectively the energy-momentum tensor of the matter and of the radiation fluid. The~expressions in \eqref{perfettiloro} indicate that matter and photons are described as two perfect fluids, so~that the stress-energy tensor of the radiation fluid is isotropic in the radiation rest-frame. 
This~is exactly the $M_1$ closure scheme described by~\cite{Sadowski2013}. 

Up to this point, we have not made any assumption about the equation of state of the radiation fluid. 

\subsection{The Dissipative Terms}\label{dissipSbup}

We have seen that the model is constructed with four currents, but the locking constraints \eqref{uytgbnm} imply that there are only two independent four-velocities. Combining this with Equation~\eqref{3newt} we find that there is only one independent four-force which needs to be provided by microphysics, which~is
\begin{equation}
G_\nu := \mathcal{R}^n_\nu + \mathcal{R}^{s_n}_\nu = -\mathcal{R}^\upgamma_\nu-\mathcal{R}^{s_\upgamma}_\nu.
\end{equation} 

Therefore, Equation \eqref{Rx} are given by
\begin{equation}\label{fghjhfd345}
\begin{split}
 G_\nu = & \,  2n^\rho \nabla_{[\rho} \mu_{\nu]} + 2s_n^\rho \nabla_{[\rho} \Theta^n_{\nu]} + \Theta^n_\nu \nabla_\rho s_n^\rho \\
  -  G_\nu = & -2\upgamma^\rho \nabla_{[\rho} \mathbb{A}^\upgamma_{\nu]}-\mathbb{A}^\upgamma_\nu \nabla_\rho \upgamma^\rho + 2s_\upgamma^\rho \nabla_{[\rho} \Theta^\upgamma_{\nu]} + \Theta^\upgamma_\nu \nabla_\rho s_\upgamma^\rho \\
\end{split}
\end{equation}
where we have used the conservation of the matter current to remove the term $\mu_\nu \nabla_\rho n^\rho$ in the first equation. The~system above may seem unfamiliar at a first sight, but with a little algebra it can be shown that it is equivalent to
\begin{equation}\label{usuale}
\begin{split}
& \nabla_\rho M\indices{^\rho _\nu} =G_\nu \\
& \nabla_\rho R\indices{^\rho _\nu} =-G_\nu \, . 
\end{split}
\end{equation} 

Therefore, the~covector $G_\nu$, representing the dissipative force of the theory, corresponds to the radiation four-force density and we have finally recovered all the basic elements of the radiation hydrodynamics~\citep{mihalas_book}.

The equations of motion given in the natural multifluid form \eqref{fghjhfd345} provide an immediate insight into the thermodynamic interpretation of $G_\nu$. Let us make the orthogonal decomposition
\begin{equation}\label{gbhnjm}
G_\nu = Q u_{n\nu} + f_\nu  \spc f_\nu u_n^\nu =0.
\end{equation}

Compared to the first equation of \eqref{fghjhfd345}, and considering that the first two terms are orthogonal to $u_n^\nu$, we obtain that
\begin{equation}\label{calormio}
 Q=\Theta^n \nabla_\rho s_n^\rho \spc
f_\nu = f_\nu^{n} + f_\nu^{s_n}  . 
\end{equation}

The first equation implies that $Q$, the~projection of $G_\nu$ parallel to $u_n^\nu$, can be interpreted as the heat exchanged or produced by the matter as a result of the interaction with the radiation fluid. 
This~equation also shows us that the rate $r_{s_n}$ does not need to be provided by microphysics, because~it must  coincide with $Q/\Theta^n$. From the second equation we see that $f_\nu$ can be seen as the part of the radiation force which tends to accelerate the fluid element. 
More directly, this can be seen by  projecting the first equation of \eqref{usuale} orthogonally to $u_n^\nu$:
\begin{equation}
(\rho + P_n) u^\rho \nabla_\rho u_{n\nu} = -(\delta\indices{^\rho _\nu} + u_n^\rho u_{n\nu})\nabla_\rho P_n + f_\nu.
\end{equation}

This~is nothing but Newton's second law for a matter fluid element, the~inertia of which is the enthalpy and which is subject to the action of a pressure force and of the radiation force $f_\nu$.

It is interesting to remark that from the thermodynamic point of view, $G_\nu$ is the local version (per unit space-time volume) of the heat four-vector acting on the matter fluid, in agreement with the covariant definition proposed by~\cite{Gavassino2020Zeroth}. Thus, $Q$ and $f_\nu$ can be rigorously identified respectively with the heat and the friction (per unit space-time volume) experienced by the matter element.

Now, let us turn our attention to the second equation of \eqref{fghjhfd345}. 
If we contract it with $u_\upgamma^\nu$ and invoke the decomposition \eqref{gbhnjm} we obtain
\begin{equation}\label{fighetto}
\Theta^\upgamma \nabla_\rho s_\upgamma^\rho = \mathbb{A}^\upgamma \nabla_\rho \upgamma^\rho + f_\nu u_\upgamma^\nu -Q \Gamma_{n\upgamma},
\end{equation}
where we have introduced the Lorentz factor
\begin{equation}
\Gamma_{n\upgamma} = -u_{n\nu}u_\upgamma^\nu.
\end{equation}

Equation \eqref{fighetto} implies that once $G_\nu$ and $r_\upgamma$ are provided by microphysics, $r_{s_\upgamma}$ is automatically constrained. Thus, our analysis indicates  that out of three reaction-type rates $r_{s_n}$, $r_{s_\upgamma}$ and $r_\upgamma$, only~one needs to be given as an external input. 
One of them, say $r_\upgamma$, should be provided by microphysics: the independent degrees of freedom of the model are 10 (i.e.,~the two scalars $s_n$, $s_\upgamma$ and the eight components $n^\nu$, $\upgamma^\nu$) but there are 5 conservation laws given by Equations \eqref{conservation} and \eqref{frftgthyh}. Hence,~there is~room for 5 equations of motion. When the functional dependence of a dissipative term is provided by means of microphysics, the~relation which defines it (i.e.,~Equation~\eqref{ratesqwe} or \eqref{Rx}) becomes an equation of motion. Therefore, we need 5 independent microphysical inputs to close the system, namely the four components of $G_\nu$ and the scalar  $r_\upgamma$. 

There is also a more physical argument to justify why microphysics should provide both the force and $r_\upgamma$: $G_\nu$ represents the energy-momentum exchange per unit time between the matter and the radiation fluid, but the same exchange may be  originated by scattering processes (which preserve the number of radiation particles) and absorption/emission processes (which modify the number of photons). This~immediately tells us that the knowledge of the force $G_\nu$ is not sufficient to constrain $r_\upgamma$. 

In the standard approach which is used in the literature (see e.g.,~\cite{Sadowski2013}) there is no need to provide $r_\upgamma$ and all the knowledge about the interaction processes between matter and radiation is incorporated into $G_\nu$. This~apparent contradiction with the multifluid approach disappears if the radiation fluid is modeled as an ideal ultrarelativistic gas. 
In fact, under this condition, the~relation
\begin{equation}\label{frgthy}
P_\upgamma = \dfrac{1}{3} \varepsilon
\end{equation}
holds not only as an equation of state, but also as a kinematic identity (i.e.,~it is valid also out of thermodynamic equilibrium).
This~implies that if there are two different thermodynamic states $(\upgamma\, , \, s_{\upgamma})$ and $(\upgamma' \, , \, s_{\upgamma}')$ such that
\begin{equation}
\varepsilon (\upgamma,s_{\upgamma}) = \varepsilon (\upgamma',s_{\upgamma}'),
\end{equation}
then they will have also the same pressure (this~is commonly enclosed in the statement that the second viscosity coefficient of an ultrarelativistic ideal gas is always identically zero~\citep{landau6,BulkGavassino}).
The mathematical implication is that if in our model the degrees of freedom of the radiation fluid are $5$ (i.e.,~$\upgamma^\nu$ and $s_\upgamma$), the~energy-momentum tensor 
\begin{equation}\label{radfluuuu}
R^{\nu \rho} = \dfrac{4}{3} \varepsilon u_\upgamma^\nu u_\upgamma^\rho + \dfrac{1}{3} \varepsilon g^{\nu \rho}.
\end{equation}
is degenerate and only 4 independent degrees of freedom have to be specified (i.e.,~$\varepsilon$ and $u_\upgamma^\nu$). 
Invoking the  expression \eqref{radfluuuu} for $R^{\nu \rho}$, the~10 degrees of freedom can be reduced to~9. 
Therefore, using the equation of motion in the form \eqref{usuale} together with the matter-particles conservation \eqref{frftgthyh}, it is possible to obtain a closed system of 9 equations, in agreement with the standard approach.

We remark, however, that in doing so one is implicitly making the assumption that also $G_\nu$ has the same degeneracy, in particular that it is not affected by deviations of $\mathbb{A}^\upgamma$ from zero. 
The conditions under which this assumption is verified will be discussed in Section~\ref{centro}. For now we will keep our analysis general, maintaining the general multifluid formulation based on 10 degrees of freedom.

\subsection{Thermodynamic Analysis of the Dissipative Terms}
\label{TermoAnalysis}

One of the biggest advantages of working in the multifluid framework (with 10 degrees of freedom) is that it keeps direct contact with the thermodynamics of the system. In this subsection we show how it can be used to derive useful thermodynamic relations which remain hidden in the standard formulation (based on 9 degrees of freedom) discussed in the previous subsection. 
These~relations can be particularly useful in those situations in which the radiation can have a finite chemical potential for a long time (like in scattering-dominated materials) in which the standard approach may be inapplicable if $G_\nu$ strongly depends on $\mathbb{A}^\upgamma$.

To provide a clear comparison with the existing literature it is convenient to work in the rest frame of the matter fluid. To do this we define a tetrad (i.e.,~an orthonormal basis of the tangent space) $e_a = e_a^\nu \partial_\nu$ which is comoving with the matter-fluid element, namely $e_0 = u_n$. 
We use this tetrad to decompose the radiation stress-energy tensor as 
\begin{equation}\label{deComPose}
\begin{split}
& R^{00} = \hat{\varepsilon} \\
& R^{0j}=R^{j0} = F^j \\
& R^{jk}=\hat{P}^{jk}\,  ,  
\end{split}
\end{equation}
which are the radiation energy density, the~radiation flux and the radiation pressure tensor (or~radiative~stress) in the matter rest-frame~\citep{mihalas_book}. By comparison with \eqref{radfluuuu}, we find that
 \begin{equation}\label{zoooooom}
\begin{split}
& \hat{\varepsilon} = (4\Gamma_{n\upgamma}^2-1)\dfrac{\varepsilon}{3} \\
&  F^j =  \dfrac{4}{3} \varepsilon \Gamma_{n\upgamma}^2 v^j \\
& \hat{P}^{jk} = (4\Gamma_{n\upgamma}^2 v^j v^k + \eta^{jk} )\dfrac{\varepsilon}{3}, \\
\end{split}
\end{equation} 
where we have introduced the three-velocity $v^j=u_\upgamma^j/u_\upgamma^0$. 
The pressure tensor $\hat{P}^{jk}$ can be written entirely in terms of $\hat{\varepsilon}$ and $F^j$: 
correcting a typo in Equation (\ref{euqation34}) of~\cite{Sadowski2013}, in~accordance with~\cite{DUBROCA1999},
we obtain 
\begin{equation}\label{pressuretensor}
\hat{P}^{jk}= \bigg( \dfrac{1-z}{2} \eta^{jk} +\dfrac{3z -1}{2 }  \dfrac{\mathcal{F}^j \mathcal{F}^k}{ \mathcal{F}^l \mathcal{F}_l} \bigg)\hat{\varepsilon} \, ,
\end{equation}
where $\mathcal{F}^j$ is the reduced radiative flux and $z$ is the Eddington factor~\citep{Levermore1984}, 
\begin{equation}
\mathcal{F}^j 
=
\dfrac{F^j}{\hat{\varepsilon}} 
\spc 
z =\dfrac{3+ 4 \, \mathcal{F}^l \mathcal{F}_l}{5+2\sqrt{4-3 \, \mathcal{F}^l \mathcal{F}_l}} \, .
\end{equation} 
Equation \eqref{pressuretensor} contains both the essence and the limitations of the closure scheme: $\hat{\varepsilon}$, $F^j$ and $\hat{P}^{jk}$ are respectively the zeroth, the~first and the second moment of the radiation specific intensity~\citep{mihalas_book} and we are closing the system by assuming that the last can be uniquely written in terms of the first two. 
This~has also the natural implication that in the reference frame of the matter element, the~isotropy is broken only along the direction identified by $F^j$. For this reason, if we use the tetrad $e_a$ to decompose the radiation four-force,
\begin{equation}
G^0=Q \spc  G^j=f^j,
\end{equation}
see Equation~\eqref{gbhnjm}, it is legitimate  to assume that
\begin{equation}\label{gringo2}
f^j \, = \, \chi \, F^j \, .
\end{equation} 

The coefficient $\chi$ can be interpreted as the total opacity, a parameter that sets the attenuation rate of the radiation flux in terms of the flux itself. If, now, we promote the three-vector $F^j$ to a space-like four-vector through the construction $F^\nu := F^j e_j^\nu$, we have the orthogonal decomposition
\begin{equation}\label{DECoMP}
G_\nu = Q \, u_{n\nu} + \chi \, F_\nu.
\end{equation}

Therefore, using simple geometrical assumptions, the~5 independent dissipative terms of the theory were reduced to 3: $Q$, $\chi$ and $r_\upgamma$. 

It is now possible to make a thermodynamic study of these terms near equilibrium. We can rewrite Equation~\eqref{dddddddri} using \eqref{calormio} and \eqref{fighetto}, obtaining  the entropy production
\begin{equation}\label{entropiaAAAA}
\nabla_\rho s^\rho = \dfrac{\mathbb{A}^\upgamma}{\Theta^\upgamma} r_\upgamma +\dfrac{\Gamma_{n\upgamma}}{\Theta^\upgamma}  F\Delta \chi +\bigg(\dfrac{1}{\Theta^n} -\dfrac{\Gamma_{n\upgamma}}{\Theta^\upgamma} \bigg)Q \geq 0,
\end{equation}
where we have introduced the relative speed $\Delta$ and the scalar $F$, defined through the relations
\begin{equation}
\Gamma_{n\upgamma} = \dfrac{1}{\sqrt{1-\Delta^2}} \spc F=\sqrt{F_\nu F^\nu}.
\end{equation}

Equation \eqref{entropiaAAAA} shows that the entropy production is the sum of three contributions. Since only the total is constrained to be non-negative, in principle far from equilibrium $r_\upgamma$, $\chi$ and $Q$ can have arbitrary sign, provided that all the contributions compensate each other giving $r_s \geq 0$. However,~if~we limit ourselves to near-equilibrium situations it is possible to obtain stronger constraints. 

Fist of all, we assume that the 3 dissipative terms $Q$, $\chi$ and $r_\upgamma$ are functions only of the local thermodynamic state of the multifluid. This~implies that in principle they are functions of 5 independent thermodynamic variables (2 identifying the thermodynamic state of the matter, 2~identifying the thermodynamic state of the radiation, 1 identifying the relative motion). In particular, we decide to work with the 5 state variables
\begin{equation}\label{chart}
(n \, , \, \Theta^n \, , \, \mathbb{A}^\upgamma \, , \, F \, , \, \Theta^\upgamma-\Theta^n) \, ,
\end{equation}
that turn out to constitute a convenient choice since it is easy to check that the local thermodynamic equilibrium state is given by
\begin{equation}\label{ellequilibrio}
\mathbb{A}^\upgamma =0  \spc F=0  \spc \Theta^\upgamma-\Theta^n = 0 \, .
\end{equation}

This~makes  $\mathbb{A}^\upgamma$, $F$ and $\Theta^\upgamma-\Theta^n$  the natural variables that can be used to parametrise a displacement of the system from local thermodynamic equilibrium. 
If we expand the dissipative terms to the linear order in these variables we obtain
\begin{equation}\label{EEEEEspandiamo}
\begin{split}
& r_\upgamma= \Xi_{\upgamma \upgamma}\mathbb{A}^\upgamma + \Xi_{\upgamma T} (\Theta^\upgamma -\Theta^n) \\
& \chi= \chi_o + \chi_\upgamma \mathbb{A}^\upgamma + \chi_T (\Theta^\upgamma -\Theta^n) \\
& Q= k_\upgamma \mathbb{A}^\upgamma+ k_T (\Theta^\upgamma -\Theta^n) \, .
\end{split}
\end{equation}

Since $G_\nu$ and $r_\upgamma$ vanish at equilibrium, there are no zeroth order terms in the above expansions for $r_\upgamma$ and $Q$. 
Moreover, there are no contributions at the first order coming from $F$ due to the symmetry of the coefficients under a transformation $F^j \longrightarrow -F^j$. The~7 expansion coefficients are all functions of the thermodynamic properties of matter only, i.e.,~$n$ and $\Theta^n$. 
Onsager's principle imposes the reciprocal relation (see Appendix \ref{appendixB} for the proof)
\begin{equation}\label{teRRmo}
k_\upgamma = \Theta^n \, \Xi_{\upgamma T}.
\end{equation}

After rewriting the entropy production Equation~\eqref{entropiaAAAA} by using the expansion \eqref{EEEEEspandiamo}, keeping only the second order in the displacement from equilibrium (which also implies $\Gamma_{n\upgamma} \approx 1$) and imposing the positivity for all small deviation from equilibrium, we obtain the conditions  
\begin{equation}\label{sonopositivo}
\Xi_{\upgamma \upgamma} \geq 0  \spc \chi_o \geq 0  \spc k_T \geq 0
\end{equation}
and, making use of the reciprocal relation \eqref{teRRmo},
\begin{equation}\label{determinant}
k_T \, \Xi_{\upgamma \upgamma} \geq \Theta^n \, \Xi_{\upgamma T}^2\, .
\end{equation}

The coefficients $\chi_\upgamma$ and $\chi_T$ appear only in higher order terms in the equation for the entropy production \eqref{entropiaAAAA}; for this reason they can be neglected in the present study. 

The constraints \eqref{teRRmo}, \eqref{sonopositivo} and \eqref{determinant} hold independently from the details of the matter-radiation interaction, so~they can be used to check the thermodynamic consistency of any model of radiation hydrodynamics.

\subsection{Application: Deriving the Four-Force $G_\nu$ from Thermodynamic Arguments}\label{centro}

The radiation four-force is usually (see e.g.,~\cite{Shapiro1996,Farris2008,rezzolla_book}) computed from the kinetic theory of radiation assuming that

\begin{enumerate}[leftmargin=8mm,labelsep=4.5mm]
\item[] i - the thermal coefficients obey the Kirchhoff law,
\item[] ii - the scattering is isotropic and coherent,
\item[] iii - the opacities have a grey-body form.
\end{enumerate}
The line of reasoning which leads to an expression for  $G_\nu$ starting from the foregoing assumptions is briefly sketched in Appendix \ref{Rilassati}. 

As a first  application of our thermodynamic study, we now show that the same form of $G_\nu$ can be derived directly in a hydrodynamic framework if one requires that
\begin{enumerate}[leftmargin=8mm,labelsep=5.5mm]
\item[] i - every dissipative process contributes additively to the transport coefficients and the thermodynamic constraints presented in the previous subsection hold separately for every microscopic contribution,
\item[] ii - the degeneracy assumption we presented in Section~\ref{dissipSbup}, which~allows reducing the degrees of freedom of the model from 10 to 9, is fulfilled also by $G_\nu$.
\end{enumerate}
Let us assume that this is the case, namely that the coefficients $\chi$ and $Q$ can be written as functions of 4 independent state variables only, instead of the 5  in \eqref{chart}. We retain the variables $n$ and $\Theta^n$ because they identify the state of the matter fluid. From  \eqref{pressuretensor} we know that the flux $F^j$ and the energy density $\hat{\varepsilon}$ can be used to identify the radiation energy-momentum tensor completely, which~in turn constitutes the reduced degree of freedom of the model. It follows that the natural choice of variables now is
\begin{equation}
(n,\Theta^n, F,\hat{\varepsilon}).
\end{equation}  

As in the previous subsection, we can perform a linear expansion in the deviations from equilibrium. As we saw, the~deviations of $\chi$ are of higher order in the model (in Equation~\eqref{decomp} $F_\nu$ is already a fist-order term), therefore we will not analyse them and we will focus our attention on $Q$. Recalling that the symmetries of the problem impose that the linear corrections in $F$ must vanish, the~expansion contains only one term, namely
\begin{equation}\label{gringo}
Q=k(\hat{\varepsilon}-\hat{\varepsilon}_{\text{eq}}).
\end{equation}

The function $\hat{\varepsilon}_{\text{eq}}$ is the equilibrium value of the energy density, which by comparison with \eqref{ellequilibrio}, is 
\begin{equation}
\hat{\varepsilon}_{\text{eq}} = \varepsilon(\Theta^\upgamma=\Theta^n,\mathbb{A}^\upgamma=0).
\end{equation}

The assumption that the radiation fluid is an ultrarelativistic ideal gas implies 
\begin{equation}
\hat{\varepsilon}_{\text{eq}} = 4\pi \hat{B}(\Theta^n) \, ,
\end{equation}
where $\hat{B}(\Theta)$ is the frequency integrated equilibrium intensity, which~can be written as 
\begin{equation}
 \hat{B}(\Theta)= \dfrac{a_R {\Theta}^4}{4\pi} \, .
\end{equation}

The coefficient $a_R$ is a constant factor which depends on the type of radiation~\citep{Farris2008}.

The expansion \eqref{gringo} is a particular case of \eqref{EEEEEspandiamo}. It is possible to relate $k$ with $k_\upgamma$ and $k_T$ considering that to first order
\begin{equation}
\hat{\varepsilon} \approx  4\pi \hat{B} -3 \upgamma \mathbb{A}^{\upgamma} + 3 s_\upgamma (\Theta^\upgamma -\Theta^n).
\end{equation}

This~expression was obtained by computing the first-order expansion coefficients of $\hat{\varepsilon}$ from the equation of state \eqref{frgthy}, together with the fact that \eqref{PRRraDD} defines the Legendre transformation 
\begin{equation}\label{difffffff}
dP_\upgamma = s_\upgamma d\Theta^\upgamma -\upgamma d\mathbb{A}^\upgamma.
\end{equation}

Plugging this expansion into \eqref{gringo} we find
\begin{equation}
Q=-3\upgamma k\mathbb{A}^{\upgamma} + 3 s_\upgamma k(\Theta^\upgamma -\Theta^n),
\end{equation}
which by comparison with the general formula \eqref{EEEEEspandiamo}, gives the relations
\begin{equation}\label{persempre}
k_\upgamma = -3 \upgamma k   \spc  k_T = 3 s_\upgamma k .
\end{equation}

We recall that since we are making a linear study, the~densities $\upgamma$ and $s_\upgamma$ can be identified with those in equilibrium, which~for an ulrarelativistic ideal gas satisfy the relation
\begin{equation}\label{contrainatoygviyg}
s_\upgamma = b_R \upgamma,
\end{equation}
where the specific entropy $b_R$ of the radiation gas is a constant ($b_R \approx 3.6$ for a Bose gas and $b_R \approx 4.2$ for a Majorana Fermi gas). 

Therefore, we have proven that the multifluid formulation of the radiation hydrodynamics, which~is a theory with 10 degrees of freedom, reduces (for small deviations from equilibrium) to the standard formulation based on 9 degrees of freedom if and only if
\begin{equation}\label{cetro2}
\dfrac{k_T}{ k_\upgamma}  = - b_R \, .
\end{equation} 

To complete the reduction to the standard theory and to see the implications of \eqref{cetro2} we divide the dissipative processes at the origin of $G_\nu$ and $r_\upgamma$ into three different categories. We call elastic scatterings ($e$) those processes which conserve the number and the total energy of the radiation particles which are involved (measured in the fluid rest-frame). 
These processes give no contribution to $r_\upgamma$ and $Q$. The~inelastic scattering processes ($I$) are those in which only the number of radiation particles is conserved. These do not give any contribution to $r_\upgamma$. Finally we have the absorption processes ($A$), which~do not conserve the radiation particle number. Note that the absorption processes include also the emission processes. In fact, absorption and emission processes are the time reversed of each other and are mediated by the same matrix element, which implies that in thermodynamic equilibrium they must obey the detailed balance. As a result, it is necessary to consider their joint action at a thermodynamic level and they must not be separated. Using our assumption (i), we can, thus, split the dissipative coefficients according to the different contributions as
\begin{equation}
\begin{split}
& r_\upgamma = r_\upgamma^A \\
& Q= Q^A + Q^I \\
& \chi = \chi^A + \chi^I +\chi^e . 
\end{split}
\end{equation} 
 
This~separation will also result into a kinetic subdivision of the expansion coefficients given in \eqref{EEEEEspandiamo} and \eqref{gringo}. 

From a purely thermodynamic point of view, the~constraints \eqref{sonopositivo} and \eqref{determinant} and the reciprocal relation \eqref{teRRmo} hold in principle only for the total coefficients, not for the separate contributions. For example we have the constraint $\chi^A + \chi^I +\chi^e \geq 0$, but this does not necessarily imply the separate non-negativity of all the three parts (because only the total appears in \eqref{entropiaAAAA}).
%However, in the computation from kinetic theory of these coefficients, the~three types of precesses are usually studied separately and treated as independent phenomena which do not affect each other. Following this approach, if one makes the thought experiment of switching off two processes (e.g. absorption and elastic scattering), this should leave the coefficients associated with the third process (in our example $Q^I$ and $\chi^I$) unchanged. On the other hand, in this hypothetical context, all the thermodynamic constraints still need to hold, and must be respected by the only process we are keeping. 
Our assumption (i), on the other hand, consists of requiring that all the thermodynamic constraints hold separately, so~that in our~example
\begin{equation}
\chi^A \geq 0 \spc \chi^I \geq 0  \spc \chi^e \geq 0, 
\end{equation}
which is in agreement with their interpretation as opacities (and, therefore, as inverses of mean-free-paths). Now, if we turn our attention to Onsager's relation \eqref{teRRmo}, for the case of the inelastic scattering we  find
\begin{equation}
k^I_\upgamma=0,
\end{equation}
which using Equation~\eqref{persempre}, gives
\begin{equation}
k^I =0 \, .
\end{equation}

We have verified that in order for \eqref{gringo} to hold, one should assume that all the scattering processes are elastic, giving $Q^I = \chi^I =0$. This~is consistent with the assumption of coherent scattering invoked by~\cite{Shapiro1996,Farris2008,rezzolla_book}.

There is a final constraint we can impose on the kinetic coefficients which arises directly from the assumption \eqref{gringo}. Let us consider a situation in which all the radiation particles are in their equilibrium distribution, but there is an excess of one particle of momentum $\vect{q}$ (measured in the rest frame of the matter). 
This~condition can be modelled as a state of the system in which
\begin{equation}
\hat{\varepsilon} =  4\pi \hat{B} +|\vect{q}|  \spc F^j = q^j.
\end{equation} 

Ignoring the scattering processes, this particle will have a life-time $\tau_A$ before being absorbed. The~absorption process can be modelled as the action of the radiation four-force for a time $\tau_A$, giving the conditions
\begin{equation}
Q\tau_A = |\vect{q}|  \spc  f\tau_A = |\vect{q}|.
\end{equation}

Simplifying the energy of the particle with the aid of \eqref{gringo} and \eqref{gringo2} we obtain $k^A\tau_A=1$ and $\chi^A\tau_A=1$, which~imply
\begin{equation}\label{cinetioco}
k^A = \chi^A.
\end{equation}

Therefore, the~multifluid approach is equivalent to the one presented by~\cite{Shapiro1996,Farris2008,rezzolla_book} if the radiation four-force can be put into the form
\begin{equation}\label{fiducia}
G^\nu = \chi^A (\hat{\varepsilon}-4\pi \hat{B})u_n^\nu + (\chi^A+\chi^e)F^\nu,
\end{equation}
which is what we wanted to prove.

%This~expression, which~is usually derived from kinetic theory under the gray-body assumption~\citep{Shapiro1996,Farris2008}, is therefore the only thermodynamically consistent way of implementing the radiation four-force in a hydrodynamic model which retains only 9 degrees of freedom.  

\subsection{Deriving the Reaction Rate From Thermodynamic Arguments}

If we assume $G_\nu$ to be given by Equation~\eqref{fiducia}, we are automatically assigning a value to $\Xi_{\upgamma T}$, as a result of the Onsager relation \eqref{teRRmo}. On the other hand, $\Xi_{\upgamma \upgamma}$ remains undetermined, therefore we are not constraining $r_\upgamma$ completely. In Appendix \ref{Rilassati}, however, we show that the same microscopic assumptions which lead to \eqref{fiducia} can be invoked to prove (assuming a non-relativistic relative speed between the matter and the radiation fluid) that 
\begin{equation}\label{kapusta}
r_\upgamma = \chi^A (\upgamma_{\text{eq}}-\upgamma),
\end{equation}
where
\begin{equation}
\upgamma_{\text{eq}} = \upgamma(\Theta^\upgamma=\Theta^n,\mathbb{A}^\upgamma=0).
\end{equation}

This~expression for the rate was adopted in the literature to model systems in which photon-conserving processes are dominant (see e.g.,~\cite{SadowskiPhotons2015}) and can be used to derive a formula for $\Xi_{\upgamma \upgamma}$. 

Before doing this, however, as a second application of our formalism, we will prove that \eqref{kapusta} is consistent with the Onsager principle. More specifically, we will show that if one assumes that the rate can be written in the generic form
\begin{equation}\label{upsilonNN}
r_\upgamma = \Upsilon(\upgamma_{\text{eq}}-\upgamma),
\end{equation}  
and $G_\nu$ is given by \eqref{fiducia}, then the Onsager relation \eqref{teRRmo} demands
\begin{equation}
\Upsilon = \chi^A.
\end{equation}

To do this, we expand $\upgamma$ near equilibrium,
\begin{equation}\label{expoarea}
\upgamma = \upgamma_{\text{eq}} + \dfrac{\partial \upgamma}{\partial \mathbb{A}^\upgamma} \mathbb{A}^\upgamma + \dfrac{\partial \upgamma}{\partial \Theta^\upgamma} (\Theta^\upgamma -\Theta^n),
\end{equation}
and compute the partial derivatives of $\upgamma$ (in equilibrium) from the ideal gas equation of state, obtaining
\begin{equation}
\dfrac{\partial \upgamma}{\partial \mathbb{A}^\upgamma} = - \dfrac{c_R \upgamma}{ \Theta^n}  \spc \dfrac{\partial \upgamma}{\partial \Theta^\upgamma} = \dfrac{3 \upgamma}{ \Theta^n},
\end{equation}
where $c_R$ is a constant coefficient ($c_R \approx 1.37$ for a Bose gas and $c_R \approx 0.91$ for a Majorana Fermi gas). The second relation can be easily obtained taking the derivative with respect to $\Theta^\upgamma$ of the (equilibrium) relation \eqref{contrainatoygviyg}. 

Employing the expansion \eqref{expoarea}, Equation~\eqref{upsilonNN} can, therefore, be rewritten as
\begin{equation}
r_\upgamma =  \dfrac{c_R \upgamma \Upsilon}{ \Theta^\n}  \mathbb{A}^\upgamma -  \dfrac{3 \upgamma \Upsilon}{ \Theta^\n} (\Theta^\upgamma -\Theta^n).
\end{equation}

Comparing with \eqref{EEEEEspandiamo} we obtain
\begin{equation}
\Xi_{\upgamma \upgamma} =  \dfrac{c_R \upgamma \Upsilon }{ \Theta^n}  \spc   \Xi_{\upgamma T} =  -  \dfrac{3 \upgamma \Upsilon}{ \Theta^n} 
\end{equation}

The Onsager relation \eqref{teRRmo}, then, implies
\begin{equation}
k_\upgamma = -3\upgamma \Upsilon,
\end{equation}
which can be compared with \eqref{persempre}. Recalling that $k=\chi^A$, we finally obtain $\Upsilon = \chi^A$, which~is what we wanted to prove.

In conclusion, we have shown that (if one negects the compton scattering) the model for radiation hydrodynamics adopted by~\cite{SadowskiPhotons2015} can be translated into the multifluid framework by~imposing
\begin{equation}
\begin{split}
& \Xi_{\upgamma \upgamma} =  \dfrac{c_R \upgamma \chi^A }{ \Theta^n}  \spc   \Xi_{\upgamma T} =  -  \dfrac{3 \upgamma \chi^A}{ \Theta^n}  \\
& k_\upgamma = -3 \upgamma \chi^A  \spc  k_T = 3 b_R \upgamma \chi^A \\
& \chi = \chi^A +\chi^e, \\
\end{split}
\end{equation}
and that this choice respects all the thermodynamic constraints we derived in Section~\ref{TermoAnalysis}. In fact, the~only non-trivial inequality which is left to check is \eqref{determinant}, which~in our case reduces to
\begin{equation}
b_R  c_R \geq 3
\end{equation}
and is satisfied by both the Bose gas ($b_R  c_R \approx 4.93$) and the Majorana Fermi gas ($b_R  c_R \approx 3.83$).

\section{Radiation as a Source of Bulk Viscosity}

In Section~\ref{HTBV} we anticipated that relativistic models for heat conduction and bulk viscosity can be naturally obtained as particular cases of the general multifluid theory. 
Hence, also any hydrodynamic model which is formulated in a multifluid framework can be interpreted as a heat-conducting  or a bulk-viscous fluid whenever it arises from a Lagrangian density of the form \eqref{zs} or \eqref{bulkio}  and there is only one strictly conserved current $n^\nu$. 
It is clear that the model for radiation hydrodynamics we presented fails to satisfy the first condition and therefore does not admit a straightforward interpretation as a heat-conducting  or as a bulk-viscous fluid. It is possible, however, to impose further constraints, besides those given in Equation~\eqref{uytgbnm}, to recover the canonical models for dissipation given in Section~\ref{HTBV}. In this subsection we focus on the possibility of transforming the presence of radiation into a contribution to bulk viscosity.

First, to recover a model for pure bulk viscosity the system should be non-conducting. 
To~implement this physical requirement, we impose the constraint
\begin{equation}\label{asdfghjkl}
n^{[\nu} \upgamma^{\rho]} =0,
\end{equation}
which implies $u_n^\nu = u_\upgamma^\nu$. This~can be obtained at a dynamic level taking the limit $\chi \longrightarrow +\infty$ and  it corresponds to the infinitely optically thick regime in which
\begin{equation}
F^\nu =0 \, .
\end{equation}

Under this condition the radiation fluid is completely advected by the matter fluid (i.e.,~there is no net conduction of photons in the matter frame). Since now matter and radiation have the same rest-frame, then it is possible to define the conglomerate internal energy
\begin{equation}
\mathcal{U} = \rho +\varepsilon,
\end{equation}
which by comparison with \eqref{perfettiloro}, is the total energy density measured in the common rest-frame. By~comparison with \eqref{LLambDDaaAqwerty} we see that the Lagrangian density has the form \eqref{bulkio}: we have constructed a bulk-viscous fluid.

An alternative way of seeing the emergence of bulk viscosity is to start from Equations \eqref{dRho}~and~\eqref{dVarepsilon} and to write 
\begin{equation}
d\mathcal{U} = \Theta^n ds_n + \mu dn +\Theta^\upgamma ds_\upgamma -\mathbb{A}^\upgamma d\upgamma,
\end{equation}
which can be recast as
\begin{equation}
d\mathcal{U} = \mu dn + \Theta^\upgamma ds  -\mathbb{A}^\upgamma d\upgamma + (\Theta^n -\Theta^\upgamma) ds_n .
\label{qwertyy}
\end{equation}

Making the identifications
\begin{equation}\label{idensirficazioni}
\Theta=\Theta^\upgamma  \spc \mathbb{A}^{s_n}= \Theta^\upgamma -\Theta^n,
\end{equation}
the differential of the internal energy \eqref{qwertyy} has the form \eqref{labulkcomelasoio}. 
% As can be seen from the equilibrium conditions \eqref{ellequilibrio}, the~displacement of the matter-radiation multifluid from local thermodynamic equilibrium is given by a non-vanishing value of the generalised affinities $\mathbb{A}^\upgamma$ and $\mathbb{A}^{s_n}$, associated with the densities $\upgamma$ and $s_n$, which~are in turn interpreted as generalised reaction coordinates.
The equilibrium conditions \eqref{ellequilibrio} tell us that the displacement of the matter-radiation multifluid from local thermodynamic equilibrium is given by a non-vanishing value of the generalised affinities $\mathbb{A}^\upgamma$ and $\mathbb{A}^{s_n}$. These affinities are  associated with the densities $\upgamma$ and $s_n$, which~are in turn interpreted as generalised reaction coordinates.
This~is in agreement with the equilibrium condition \eqref{formuloio} for bulk-viscous fluids.

The pressure and energy-momentum tensor, which~for the matter-radiation multifluid are equal to \eqref{kopf} and \eqref{TTT}, coincide with the ones calculated according to the prescription \eqref{pressandenem}, namely
\begin{equation}
\begin{split}
& \Psi = -\mathcal{U} +\mu n + \Theta s-\mathbb{A}^\upgamma \upgamma - \mathbb{A}^{s_n} s_n \\
& T^{\nu \rho} = \Psi g^{\nu \rho} + (\mathcal{U}+\Psi)u_n^\nu u_n^\rho .  
\end{split}
\end{equation}

This~is due to the fact that the variationally defined energy-momentum tensor \eqref{Tnuro} is invariant under changes of chemical basis~\citep{Carter_starting_point,Termo}.

We can also map the dissipative expansion coefficients introduced in \eqref{EEEEEspandiamo} into the reaction matrix $\Xi_{AB}$ presented in Equation~\eqref{uiop}. With the aid of Equation~\eqref{calormio}, we can rewrite \eqref{EEEEEspandiamo} in the form 
\begin{equation}
\begin{pmatrix}
r_\upgamma \\
r_{s_n} \\
\end{pmatrix}
=
\begin{bmatrix}
   \Xi_{\upgamma \upgamma} & \Xi_{\upgamma T}  \\
    k_\upgamma/\Theta^n & k_T/\Theta^n  \\
\end{bmatrix} 
\begin{pmatrix}
\mathbb{A}^\upgamma  \\
\mathbb{A}^{s_n} \\
\end{pmatrix},
\end{equation}
which is the defining relation for the $2\times 2$ matrix $\Xi_{AB}$. The~Onsager reciprocal relation \eqref{teRRmo} ensures the symmetry condition $\Xi_{AB}=\Xi_{BA}$, while the thermodynamic constraints \eqref{sonopositivo} and \eqref{determinant} imply the non-negativity of $\Xi_{AB}$.

We can also compute the bulk viscosity coefficient \eqref{ZZZeTa} associated with the radiation processes (the details of the calculations are reported in Appendix \ref{qazsedc}), that turns out to be
\begin{equation}
\zeta= \dfrac{k_T +2b_R\Theta^n \Xi_{\upgamma T}+b_R^2 \Theta^n \Xi_{\upgamma \upgamma} }{\Xi_{\upgamma \upgamma}k_T - \Theta^n\Xi_{\upgamma T}^2} \bigg( \dfrac{\partial x^{\text{eq}}_\upgamma}{\partial v} \bigg|_{x_s} \bigg)^2.
\end{equation}

The above formula is the general expression for the bulk viscosity of the multifluid for arbitrary values of the kinetic coefficients $k_T$, $\Xi_{\upgamma T}$ and $\Xi_{\upgamma \upgamma}$. If we impose the validity of the condition \eqref{cetro2}, which~combined with the Onsager relation \eqref{teRRmo} implies
\begin{equation}
k_T = -b_R \Theta^n \Xi_{\upgamma T},
\end{equation}
together with the consequent expression for the four-force \eqref{fiducia}, we find that the bulk viscosity coefficient simplifies to
\begin{equation}\label{fvfbgnhm}
\zeta =\dfrac{b_R \Theta^n}{3\upgamma \chi^A} \bigg( \dfrac{\partial x^{\text{eq}}_\upgamma}{\partial v} \bigg|_{x_s} \bigg)^2.
\end{equation} 

The coefficient $\Xi_{\upgamma \upgamma}$ simplifies and does not play any role in the final expression for the bulk viscosity. This~stems from the fact that the condition \eqref{cetro2} is imposed to guarantee that the hydrodynamic evolution is decoupled from the chemical evolution of the degree of freedom $\upgamma$. The~term in the brackets in Equation~\eqref{fvfbgnhm} strongly depends on the equation of state of the matter fluid {{and essentially nothing can be said a priori if} $\rho(n,s_n)$ {is not specified. In Appendix} \ref{loabbiamofatto} {we compute} $\zeta$ {explicitly for the case of a non-degenerate gas in the ultra-relativistic and non-relativistic limits.}}

%mdpi: please check if this should be a part of one paragraph, if so, please merge it into that paragraph.
%authors: this is a typo, we deleted it altogether.
%%% % % \hl{vanishes in the case in which the latter is an ultra-relativistic ideal gas.}

Equation \eqref{fvfbgnhm} can be used to replace the multifluid with a Navier-Stokes (or Israel-Stweart) bulk-viscous fluid (the final equations governing these fluids can be taken to be the ones presented in section IV-C or VII of~\cite{BulkGavassino}) using the matter+radiation equilibrium equation of state and $\zeta$ as the prescription for the bulk-viscosity coefficient.

\section{Radiation Hydrodynamics as a Model for Heat Conduction}

To obtain a model for heat conduction we need to consider an opposite situation with respect to the bulk-viscous case. We need to assume that the radiation-particle production rate and the exchange of energy in the rest-frame of the matter are faster than the hydrodynamic time-scale. 
In this way they are always in equilibrium and do not contribute to the entropy production. This, compared with \eqref{entropiaAAAA}, implies (assuming that the relative speed is non-relativistic)
\begin{equation}\label{gkl}
\Theta^n= \Theta^\upgamma \spc \mathbb{A}^\upgamma=0.
\end{equation}

This~condition can be formally achieved by sending at least two out of the three coefficients $\Xi_{\upgamma \upgamma},\Xi_{\upgamma T}$ and $k_T$ to infinity. 
In the case in which the radiation four-force has the form \eqref{fiducia}, the~limit $\Xi_{\upgamma \upgamma} \longrightarrow +\infty$ can be safely imposed (even in those cases in which this is not a rigorous assumption) because the value of $\Xi_{\upgamma \upgamma}$ does not have any influence on the hydrodynamic evolution of the model based on 9 degrees of freedom ($n^\nu,\Theta^n,\hat{\varepsilon},F^\nu$), which~are usually the variables of physical interest. 

For the other coefficients there is the complication that the constraint \eqref{cinetioco} implies that if \mbox{$k \longrightarrow +\infty$}, then also $\chi$ will diverge. However, in this case we would have also the locking condition \eqref{asdfghjkl} and the multifluid would simply reduce to a single perfect fluid. Since we need to keep $\chi$ finite to enable heat conduction, in the following we will simply assume that \eqref{gkl} holds as a mathematical constraint, without specifying under which physical conditions this constraint is respected.

\subsection{The Origin of the Entrainment}

An easy way of studying the implications of the constraint \eqref{gkl} is to analyse its effect on the Lagrangian density. From \eqref{vbvbvb} and \eqref{moMenti} we have that the variation of $\Lambda$ under arbitrary variations of the currents (at constant metric components) is
\begin{equation}
\delta \Lambda = \Theta^n_\nu \delta s_n^\nu + \mu_\nu \delta n^\nu +\Theta^\upgamma_\nu \delta s_\upgamma^\nu -\mathbb{A}^\upgamma_\nu \delta \upgamma^\nu.
\end{equation}

We can use \eqref{dfgtyh} to rewrite this variation in the equivalent form
\begin{equation}
\delta \Lambda = \Theta_\nu \delta s^\nu + \mu_\nu \delta n^\nu-\mathbb{A}^{s_n}_\nu \delta s_n^\nu  -\mathbb{A}^\upgamma_\nu \delta \upgamma^\nu,
\end{equation}
where analogously with the notation \eqref{idensirficazioni}, we have introduced the covectors
\begin{equation}\label{thetgam}
\Theta_\nu = \Theta^\upgamma_\nu 
\spc 
\mathbb{A}^{s_n}_\nu =\Theta^\upgamma_\nu - \Theta^n_\nu.
\end{equation}

Now we impose that the variations of the currents, which~in principle may be all independent, satisfy the constraints \eqref{uytgbnm}: we can write
\begin{equation}\label{loop}
\begin{split}
&  s_n^\nu = x_{sn} n^\nu  \\
&  \upgamma^\nu =y \, s_\upgamma^\nu = y (s^\nu -x_{sn}n^\nu) 
\end{split}
\end{equation}
and a variation of the first expression reads 
%(we do not need to write explicitly the variation of the second equation because we are not going to employ it, but we reported the constraint for completeness),
\begin{equation}
\delta s_n^\nu = n^\nu \delta x_{sn} + x_{sn} \delta n^\nu.
\end{equation}

The variation of $\Lambda$ becomes
\begin{equation}\label{dLambIo}
 \delta \Lambda =  \Theta_\nu \delta s^\nu + \tilde{\mu}_\nu \delta n^\nu  - \mathbb{A}^{s_n}_\nu n^\nu \delta x_{sn}-\mathbb{A}^\upgamma_\nu \delta \upgamma^\nu,
\end{equation}
where we have introduced the momentum
\begin{equation}
\tilde{\mu}_\nu = \mu_\nu-x_{sn}\mathbb{A}^{s_n}_\nu.
\end{equation}

Finally, we impose the constraints \eqref{gkl} on the non-varied state (i.e.,~the original state in which the system is, before we make the variation), but not on the varied state. 
The meaning of this procedure is that we are assuming a reference state which is solution of Equation \eqref{gkl}, but the variations $\delta s^\nu$, $\delta n^\nu$, $\delta x_{sn}$ and $\delta y$ are completely arbitrary (the reason why we leave them arbitrary will be clarified in the next subsection). It is easy to see that for non-relativistic relative speeds \eqref{gkl} is equivalent to the chemical-type equilibrium conditions
\begin{equation}\label{equilibriochimico5}
\mathbb{A}^{s_n}_\nu n^\nu=0  \spc \mathbb{A}^\upgamma_\nu =0,
\end{equation}
so that all the contributions arising from the arbitrary variations $\delta x_{sn}$ and $\delta y$ vanish and \eqref{dLambIo} reduces~to 
\begin{equation}\label{sbuf}
\delta \Lambda = \Theta_\nu \delta s^\nu + \tilde{\mu}_\nu \delta n^\nu.
\end{equation}

This~proves that the variation of the Lagrangian density is indistinguishable from the one of a heat-conducting fluid, provided that we interpret the momenta $\Theta_\nu$ and $\tilde{\mu}_\nu$ as the conjugate momenta respectively to the entropy and particle current (cfr. with Section~\ref{HCX1}). These momenta, written in terms of the currents $s^\nu$ and $n^\nu$, are respectively
\begin{equation}\label{MomMMM}
\begin{split}
& \Theta_\nu = \dfrac{\Theta^\upgamma}{s_\upgamma} (s_\nu-x_{sn}n_\nu) \\
& \tilde{\mu}_\nu = \bigg( \dfrac{\mu}{n} +x_{sn}\dfrac{\Theta^n }{n} + x_{sn}^2 \dfrac{\Theta^\upgamma}{s_\upgamma}\bigg)n^\nu - x_{sn}\dfrac{\Theta^\upgamma}{s_\upgamma} s^\nu , \\
\end{split}
\end{equation}
 which lead to the identification of the bulk coefficients
 \begin{equation}\label{LientrainoIo}
 \mathcal{B}=  \dfrac{\mu}{n} +x_{sn}\dfrac{\Theta^n }{n} + x_{sn}^2 \dfrac{\Theta^\upgamma}{s_\upgamma}  \spc \mathcal{C}= \dfrac{\Theta^\upgamma}{s_\upgamma}
 \end{equation}
and of the anomalous coefficient $\mathcal{A}^{sn}$, which~encodes the entrainment phenomenon, see Equation~\eqref{vbvbvb},
\begin{equation}
\mathcal{A}^{sn} = \mathcal{A}^{ns} = -x_{sn} \dfrac{\Theta^\upgamma}{s_\upgamma}.
\label{sicarioprezzolato}
\end{equation}

It is useful, now, to summarize what we have obtained so far. 
We started with the model \eqref{LLambDDaaAqwerty} which was built considering 4 currents. 
No entrainment was assumed in this model, namely the conjugate momenta to all the currents were collinear with the respective currents. Then we have reduced the dynamical degrees of freedom to 2 independent currents invoking 2 collinearity constraints \eqref{uytgbnm} and 2 chemical-type equilibrium conditions \eqref{equilibriochimico5}. We have found that the reduced theory, described using only 2 independent four-currents, reproduces a fluid with entrainment (i.e.,~a fluid in which the  conjugate momenta   are linear combinations of both the currents, see Section~\ref{polj}). 

This~gives us a deep insight on the fundamental nature of the entrainment. In fact, the~anomalous coefficient \eqref{sicarioprezzolato} is proportional to $x_{sn}$, which~represents  the entropy per-particle  carried by the matter fluid (see \eqref{loop}), a quantity that comoves with $n^\nu$. Thus, we see that the entrainment between two currents may arise also in a theory which originally does not admit it: an effective entrainment coupling emerges whenever a fraction of the constituents of one current is forced to comove with the second current and the processes which tend to alter this fraction are in equilibrium. 

As a final remark, we note that for the case of the superfluid Helium, this mechanism for the emergence of entrainment is at the origin of the equivalence between the Tisza-Landau two-fluid model and the multifluid model of~\cite{carter92}. In the Landau model it is assumed that the Helium current can be split into two non-conserved and non-entrained currents, one of which (the so-called normal current) is locked with the entropy current. 
On the other hand, in model of Carter and Khalatnikov  this splitting is not explicitly done, but its existence is reflected into a non-vanishing entrainment between the entropy and the total particle current. The~steps of the proof of the equivalence between these two approaches are analogous to the calculations performed in this subsection.

\subsection{Energy-Momentum Tensor}

The energy-momentum tensor introduced in Section~\ref{polj} can be equivalently defined as 
\begin{equation}\label{tesnrofjg}
T^{\nu \rho} = \dfrac{2}{\sqrt{-g}} \dfrac{\delta (\sqrt{-g}\Lambda)}{\delta g_{\nu \rho}} \bigg|_{\star n_x}  
\end{equation}
where the variation is performed keeping constant the components of the Hodge duals of the currents
\begin{equation}
(\star n_x)_{\nu \rho \sigma} =  \varepsilon_{\lambda \nu \rho \sigma } n_x^\lambda
\, .
\end{equation}

In Section~\ref{RHfoundation} the energy-momentum tensor was computed treating the currents $s_n^\nu$ and $\upgamma^\nu$ as independent variables, therefore in the calculation of the derivative \eqref{tesnrofjg} their Hodge duals where held fixed, imposing the condition
\begin{equation}
\delta x_{sn}=\delta y =0 \, .
\end{equation}

On the other hand, if we want to eliminate these currents from the set of possible degrees of freedom and work with a theory in which the only two fundamental currents are $n^\nu$ and $s^\nu$, we have to impose the constraints \eqref{gkl} also in the varied state. This~implies that $x_{sn}$ has to be considered a function of the three fundamental scalars of the model, giving 
\begin{equation}
\delta x_{sn} = \dfrac{\partial x_{sn}}{\partial n^2} \delta (n^2)+\dfrac{\partial x_{sn}}{\partial s^2} \delta (s^2)+\dfrac{\partial x_{sn}}{\partial n_{ns}^2} \delta (n_{ns}^2).
\end{equation}
The same argument should in principle hold also for $y$; however we know from microphysics that in equilibrium $y=1/b_R$, therefore the condition $\delta y=0$ is left unchanged.

However, in deriving the formula of the variation \eqref{sbuf} no assumption on the variation of $x_{sn}$ and $y$ was made (they were completely arbitrary). This~means that the energy-momentum tensor we obtain from the formula \eqref{tesnrofjg} is the same both in the original model with 4 currents, imposing the constraints only at a dynamical level, and in the reduced model with 2 currents, in which the constraints hold also off-shell (and therefore remain valid when the variation is performed, see Section 4 of~\cite{carter92}).
The~implication is that the pressure \eqref{kopf} and energy-momentum tensor \eqref{TTT} can be equivalently rewritten from \eqref{sbuf} in the canonical forms \eqref{preSSSSSSSSSSSSSSSSSSSSSUUUUUUUUUUre} and \eqref{Tnuro}:
\begin{equation}
\begin{split}
& \Psi = \Lambda - n^\nu \tilde{\mu}_\nu -s^\nu \Theta_\nu \\
& T \indices{^\nu _\rho} =\Psi \delta\indices{^\nu _\rho}  +n^\nu \tilde{\mu}_\rho + s^\nu \Theta_\rho. \\
\end{split}
\end{equation}
{This~can also be  checked with direct calculations. Thus, we have proven that the energy-momentum tensor of the radiation hydrodynamics assumes the form of the energy-momentum tensor of a heat-conducting fluid, Equation~\eqref{decomp}, provided that we interpret the entrained momenta \eqref{MomMMM} as the canonical conjugate momenta  to the entropy current and to matter-particle current~respectively.}

{Let us perform the Eckart-frame decomposition \eqref{Eckart}. We note that the Eckart four-velocity identifies the matter rest-frame. However, the decomposition of the radiation stress-energy tensor has already been performed in Section~\ref{TermoAnalysis}. Comparing \eqref{Eckart} with \eqref{pressuretensor} we have the straightforward~identifications}
\begin{equation}
\mathcal{U} = \rho +\hat{\varepsilon} \spc  q^\nu = F^\nu,
\end{equation}
so we see that Eckart's  heat-flux is simply the radiation flux. In Appendix \ref{perchenolinonso} we show that this identification (which we have obtained from the comparison of the energy-momentum tensors) is also consistent with the Eckart-frame decomposition \eqref{entrcurr} of the entropy current \eqref{dfgtyh}.

Finally, we can study the second order term in \eqref{Eckart}, whose coefficient $\mathcal{D}$, introduced in \eqref{DDD}, is~given~by
\begin{equation}
\mathcal{D} = \dfrac{1}{\Gamma_{n\upgamma}^2 s_\upgamma \Theta^\upgamma},
\end{equation}
where we have used the second equation of \eqref{LientrainoIo}. It is easy to prove that the pressure tensor $\hat{P}^{jk}$ introduced in Section \ref{TermoAnalysis} can be written as
%mdpi: please confirm this change.
%authors: OK
\begin{equation} 
\hat{P}^{jk}= P_\upgamma  \eta^{jk} + \mathcal{D}q^j q^k,
\end{equation}
which provides, in the case of radiation hydrodynamics, an immediate microscopic interpretation of the phenomenological second order term appearing in \eqref{Eckart} as the anisotropic contribution to the radiation pressure tensor. 

\subsection{The Hydrodynamic Equations}
\label{theHYDRA}

The equations of motion \eqref{fghjhfd345} describe the dissipative interaction between the currents $n^\nu,s_n^\nu$ and the currents $\upgamma^\nu,s_\upgamma^\nu$. To complete the construction of the model for heat conduction we need to recast these equations into the form \eqref{jujujuj}, which~describes the dissipative interaction between $n^\nu$ and $s^\nu$. 

Before doing this, however, there is an important remark to make, which~was pointed out in~\cite{Carter_starting_point}. Let us consider a generic multifluid and define the forces
\begin{equation}\label{Ryx}
\mathcal{R}\indices{_x ^y _\nu} = 2n_x^\rho \nabla_{[\rho} \mu^y_{\nu]} + \mu_\nu^y \nabla_\rho n_x^\rho.
\end{equation}

They might be thought to constitute a $(1,1)$ tensor in the chemical species index $x$. In fact, if we change the fundamental currents of our theory through a change of chemical basis, i.e.,~a transformation 
\begin{equation}
\tilde{n}^\nu_x = N\indices{_x^y}n_y^\nu,
\end{equation}
where the coefficients $N\indices{_x^y}$ are some constants, then the conjugate momenta transform according to the contravariant law
\begin{equation}
\mu^y_\nu= \tilde{\mu}^x_\nu N\indices{_x^y}
\end{equation}  
and the forces \eqref{Ryx} will consequently have a mixed transformation law. Since, from \eqref{Rx}, we have that
\begin{equation}
\mathcal{R}^x_\nu = \mathcal{R}\indices{_x ^x _\nu},
\end{equation}
the dissipative forces represent only the diagonal part of the tensor $\mathcal{R}\indices{_x ^y _\nu}$. As a result, after a change of basis the new forces $\tilde{\mathcal{R}}^x_\nu$ will not be linear combinations of the old forces $\mathcal{R}^x_\nu$ only, but the summation will involve also off diagonal terms $\mathcal{R}\indices{_x ^y _\nu}$ with $x \neq y$. This~has two remarkable consequences. 

The first consequence is that if we impose  $\mathcal{R}^x_\nu =0$ for all $x$  in the non-dissipative limit, in principle this will cease to hold if we change the chemical basis ($\tilde{\mathcal{R}}^x_\nu \neq 0$), due to the presence of mixed terms $\mathcal{R}\indices{_x ^y _\nu} \neq 0$, for $x \neq y$. This~shows that in general, there is no way to guarantee that the forces $\mathcal{R}^x_\nu$ vanish in a non-dissipative theory, unless one has a microscopic argument to support the choice of a preferred chemical basis in which this occurs. 
This~is   related to the problem we presented at the end of Section~\ref{InclDissip12344567889}, namely the fact that one can impose the vanishing dissipation condition \eqref{vanishDissipPa} even in a context in which $\mathcal{R}^x_\nu \neq 0$.

The second consequence is that since the terms $\mathcal{R}\indices{_x ^y _\nu}$ contain derivatives (both in space and time), if one imposes that the forces $\mathcal{R}^x_\nu$ depend only on the value of the hydrodynamic fields in the point and not on their derivatives, this will be no longer true when we change chemical basis. Therefore,~as~we have already pointed our in Section~\ref{HCX1}, one cannot impose that the forces do not depend on the derivatives of the hydrodynamic fields without a microscopic justification.

Now, we can study the heat-conductive limit of radiation hydrodynamics, knowing that
both   these complications may arise (in fact, we are going to perform  substantially a change of chemical~basis). 

Let us take the second equation of \eqref{fghjhfd345} and use the second constraint of \eqref{gkl} to remove the terms proportional to $\mathbb{A}^\upgamma_\nu$:
\begin{equation}\label{Blumfghjhfd345}
 -G_\nu =  2s_\upgamma^\rho \nabla_{[\rho} \Theta^\upgamma_{\nu]} + \Theta^\upgamma_\nu \nabla_\rho s_\upgamma^\rho . 
\end{equation}

From the second equation of \eqref{jujujuj}, we know that
\begin{equation}
\mathcal{R}^s_\nu = 2s^\rho \nabla_{[\rho} \Theta_{\nu]} + \Theta_\nu \nabla_\rho s^\rho.
\end{equation}

Recalling the first definition of \eqref{thetgam}, we can use \eqref{Blumfghjhfd345} to rewrite $\mathcal{R}^s_\nu$ in the form
\begin{equation}
\mathcal{R}^s_\nu = 2s_n^\rho \nabla_{[\rho} \Theta_{\nu]} + \Theta_\nu \nabla_\rho s_n^\rho - G_\nu.
\end{equation}

We can further simplify this expression by invoking the decomposition \eqref{DECoMP}, which recalling the second equation of \eqref{thetgam}, gives us the final formula
\begin{equation}\label{Eccoloquiwdindccwe}
\mathcal{R}^s_\nu = 2s_n^\rho \nabla_{[\rho} \Theta_{\nu]} + \mathbb{A}^{s_n}_\nu \nabla_\rho s_n^\rho - \chi F_\nu.
\end{equation}

The equation for the entropy production \eqref{entropiaAAAA}, on the other hand, reduces in our case to
\begin{equation}\label{entropiaCCCCV}
\nabla_\rho s^\rho = \dfrac{F\Delta \chi}{\Theta^\upgamma}    \geq 0.
\end{equation}

Therefore, we  see that if $\chi =0$ the theory is non-dissipative, even if $\mathcal{R}^s_\nu \neq 0$. Furthermore we note that having imposed that $\chi F_\nu$ depends only on the value of the hydrodynamic variables in the point, the~force $\mathcal{R}^s_\nu$ contains terms which are linear in the derivatives. We also note that Equation~\eqref{Eccoloquiwdindccwe} describes a force which is not necessarily a linear combination of $n^\nu$ and $s^\nu$, but contains a component which is orthogonal to both. This~is in agreement with the discussion in~\cite{AC15}. 

%Nils have studied the problem of heat conduction in a general perspective and they concluded that in principle one cannot exclude that the force $\mathcal{R}^s_\nu$ might have a contribution which is orthogonal to both the currents $n^\nu$ and $s^\nu$. For the case of radiation transport, we have  the derived Equation~\eqref{Eccoloquiwdindccwe} contains also these contributions. 

\subsection{Heat Conductivity Coefficient}

We conclude the section by calculating the coefficient of thermal conductivity. This~is conveniently done by comparing the formulas for the entropy production \eqref{kufkuf} and \eqref{entropiaCCCCV}. Imposing the equality of the two, we  find the condition (neglecting the Lorentz factors)
\begin{equation}\label{DDDelta}
\Delta \chi = \dfrac{F}{\kappa \Theta^\upgamma}.
\end{equation}

However, from \eqref{1234567} and \eqref{9876543} we obtain
\begin{equation}
F = \Theta^\upgamma s_\upgamma \Delta,
\end{equation}
which plugged into \eqref{DDDelta}, gives
\begin{equation}\label{Kappa}
\kappa = \dfrac{s_\upgamma}{\chi} =\dfrac{4}{3} \dfrac{a_R \Theta^3}{\chi^A+\chi^e} .
\end{equation}

It is well known that the radiation hydrodynamics have a diffusion-type limit which makes it analogous to a phenomenon of heat conduction. \cite{Shapiro1989} and \cite{Farris2008} have computed explicitly the corresponding coefficient $\kappa$, obtaining the formula \eqref{Kappa}.  We have generalized this result, showing the complete formal analogy between the two systems in the framework of the multifluid formalism.

\section{Limitations of the Model}

We conclude with a few comments about the limitations of our model.
We have seen that assuming the Lagrangian density \eqref{LLambDDaaAqwerty} as a starting point for Carter's approach implies that the matter and the photons  are both described  as  two  perfect  fluids.
The fact that the matter can be modelled as a perfect fluid is justified if the collisions between matter-particles are faster than the hydrodynamic time-scale. However, the~same argument cannot be applied to the radiation fluid, whose particles typically do not interact with each other {{and, therefore, an H-theorem for the radiation gas alone does not exist. This~implies that the closure scheme cannot, in general, be justified using thermodynamic or kinetic arguments. Indeed, it is well-known that the local properties of the radiation stress-energy tensor depend on the global structure of the radiation field (in particular on the disposition of its sources) and, as a consequence, the~$M_1$ closure scheme fails the multiple-source shadow test} \citep{Sadowski2013}.}

Given the fact that it is not possible to justify the closure scheme as a physical limit (and therefore it is not guaranteed to provide an accurate description of reality), we can understand its appearance in the multifluid theory as a byproduct of applying the principles of information theory in the context of Carter's formalism. In fact, one is required to provide a limited set of macroscopic parameters (the radiation particle total current $\upgamma^\nu$ and the rest-frame energy density $\varepsilon$, both appearing in the Lagrangian density \eqref{LLambDDaaAqwerty}{) and all the remaining properties of the radiation field need to be written in terms of this limited (local) information. Then, following the philosophy of information theory, well summarised by} \cite{Jaynes1}, we have to assume that the system is in the microstate that maximizes the entropy (or, equivalently, minimizes the information) compatibly with the  values of the macroscopic parameters which are known.  Thus, denoting the microscopic single-particle-state occupation numbers by $N(\vect{p})$,
the Shannon entropy $s_\upgamma$, which~for an ideal gas is the opposite of Boltzmann's H-function, is given in the radiation rest-frame by
\begin{equation}
s_\upgamma = -\int \bigg[ N \log N + j (1-j N )\log (1-j N)  \bigg] \dfrac{g \, d_3 p}{h_p^3} ,
\end{equation}
where $j=-1$ for Bosonic radiation and $+1$ for Fermionic radiation, $g$ is the spin degeneracy and $h_p$ is Planck's constant. The~particle and energy density are 
\begin{equation}
\upgamma = \int N \, \dfrac{g \, d_3 p}{h_p^3}  
\spc 
\varepsilon= \int N |\vect{p}| \, \dfrac{g\,  d_3 p}{h_p^3} .
\end{equation}

Hence, the~most probable state must be obtained imposing
\begin{equation}
\delta s_\upgamma + \alpha \, \delta \upgamma - \beta \, \delta \varepsilon =0,
\end{equation}
where $\alpha$ and $\beta$ are two Lagrange multipliers. This~operation gives the Bose-Einstein/Fermi-Dirac~occupation
\begin{equation}
N(\vect{p}) = \dfrac{1}{e^{-\alpha + \beta |\vect{p}|}+j} ,
\end{equation}
which justifies the interpretation of $s_\upgamma$, $\Theta^\upgamma = \beta^{-1}$ and $-\mathbb{A}^\upgamma = \alpha \beta^{-1}$ as respectively the thermodynamic entropy, temperature and chemical potential of the radiation gas. 

{{To summarize, the~multifluid formalism forces us to assume that the currents and the Lagrangian density are the only information we are given (it is our macroscopic knowledge about the system), and this leads to a fluid-model of the radiation gas. }}

{{Apart from failing the multiple-source shadow test,}} this approach has also clear limitations when the opacity has a strong dependence on the frequency. In fact, under this condition, the~expression \eqref{fiducia} for the radiation four-force ceases to hold and a hydrodynamic approach may be inapplicable. In this situation one should follow  the evolution of each radiation frequency separately, requiring a kinetic theory approach.

Furthermore, even in the case in which there was only one relevant radiation frequency, which~in principle may still allow the use of a hydrodynamic treatment, the~$M_1$ closure scheme of~\cite{Sadowski2013} would {{be inconsistent with the given information.}} This~was shown in the context of information theory by~\cite{Minerbo1978}, who worked in the rest frame of matter and considered  a monochromatic radiation flux with a given frequency (measured in the matter's frame).  
By choosing the energy $\hat{\varepsilon}$ and the components radiation flux $F^j$ as basic information about the system, he found a maximum entropy distribution which produces an energy-momentum tensor which obeys to a different closure scheme (see also~\cite{Levermore1984} for a comparison between the different closure schemes).

\section{Conclusions}

We have studied how radiation hydrodynamics can be modelled in the context of Carter's multifluid formalism. The~radiation stress-energy tensor was found to obey to the $M_1$ closure scheme presented by~\cite{Sadowski2013} and the hydrodynamic equations were shown to be equivalent to those which are often employed in the literature~\citep{mihalas_book}. Moreover, we connected the hydrodynamic theory with non-equilibrium thermodynamics and performed an Onsager analysis of the dissipative terms of the model. 

As an immediate application, we showed that the grey-body radiation four-force~\citep{Shapiro1996,Farris2008,rezzolla_book} is the only thermodynamically consistent expression for the force between the matter and the radiation fluid which can be used in a model with 9 independent degrees of freedom.

The multifluid formalism, therefore, perfectly captures and describes the physics of radiation hydrodynamics in detail, offering novel insight into a well understood subject.

In the second part of the paper, we reinterpreted radiation hydrodynamics as a particular case of relativistic dissipation and we used this reformulation to gain new understanding in the latter. In the infinitely optically thick limit, the~interaction between the matter and the radiation fluid was shown to become a source for bulk viscosity. This~is in accordance with the more general result  that any locally isotropic fluid is a Carter bulk-viscous multifluid~\citep{BulkGavassino}.

In the opposite limit, in which the radiation fluid is assumed to be in chemical equilibrium with respect to particle production processes and to have a rest-frame temperature equal to the one of the matter fluid, the~multifluid reduces to a model for heat conduction. We found that the entrainment between the entropy and the matter current arises naturally from the original splitting of the entropy current into a matter part and a radiation part which were not entrained. When we used the condition of equal temperature to reduce the number of degrees of freedom of the model, the~momentum associated with the entropy of matter was naturally divided into two parts which were redistributed between the currents, generating an effective entrainment in a theory that originally (because of the assumed form \eqref{LLambDDaaAqwerty} of the Lagrangian density) was entrainment-free.

The equations of motion of the resulting heat-conducting fluid have, in the parabolic limit, the~well known diffusive form of the radiative transport~\citep{Shapiro1989,Farris2008} and we verified the correspondence of the respective transport coefficients. In the hyperbolic regime, however, beyond first order in the deviations from equilibrium, the~model was shown to be different from all the proposed universal models for heat conduction~\citep{Israel_Stewart_1979,noto_rel,lopez2011}. In fact, the~structure of the hydrodynamic equations (determined by the expression of the four-force acting between the currents) preserves many details of the physics of the radiation hydrodynamics and no universal available theory for the heat conduction is so general to be able to encompass all these details. 

{{This~paper constitutes a further step forward towards the global unification of the relativistic hydrodynamics, showing with a concrete example how multifluids and dissipative single-fluids can arise as two different mathematical descriptions of the same theory. In particular, our study might constitute the beginning of the construction of a bridge between the hydrodynamic models employed in simulations of super-novae (where a multifluid approach is usually adopted) and those employed in simulations of neutron-star mergers (where a single-fluid approach is preferred).}}

\section*{Acknowledgements}

We acknowledge support from the Polish National Science Centre grants SONATA BIS 2015/18/E/ST9/00577 and OPUS 2019/33/B/ST9/00942. Partial support comes from PHAROS, COST Action CA16214. We thank Giovanni Camelio for valuable discussion.

\appendix

\section{Triangular vs. Square Formulation}\label{AAA}

Consider for simplicity the case with two currents, $n_A^\nu$ and $n_B^\nu$. According to \eqref{gggg}, the~Lagrangian density has the form
\begin{equation}\label{tttttt}
\Lambda = \Lambda (n_{AA}^2,n_{AB}^2,n_{BB}^2),
\end{equation}
whose differential, recalling \eqref{Bb} and \eqref{Aa}, is
\begin{equation}\label{tttttttt3}
d\Lambda = -\dfrac{\mathcal{B}^A}{2} d(n_{AA}^2) - \mathcal{A}^{AB} \, d(n_{AB}^2)-\dfrac{\mathcal{B}^B}{2} d(n_{BB}^2).
\end{equation}

This~representation of the equation of state gives $\Lambda$ as a function of the upper triangle $x \leq y$ of the matrix $n_{xy}^2$ and for this reason it may be called \textit{{triangular formulation}}. An other approach consists of considering $\Lambda$ as a function of the whole square matrix (and for this reason we can call it \textit{{square formulation}}) trough the equation
%mdpi: is the italic necessary?
%authors: YES
\begin{equation}\label{ttttttt2}
\Lambda(n_{AA}^2,n_{AB}^2,n_{BA}^2,n_{BB}^2) := \Lambda\bigg( n_{AA}^2, \dfrac{n_{AB}^2 + n_{BA}^2}{2},n_{BB}^2 \bigg),
\end{equation} 
where in the right-hand side we are using the functional dependence of $\Lambda$ presented in Equation~\eqref{tttttt}. Equations \eqref{tttttt} and \eqref{ttttttt2} describe the same physical quantity, in fact in any real state $n_{AB}^2 = n_{BA}^2$. However in the square formulation $n_{AB}^2$ and $n_{BA}^2$ are treated in the equation of state as independent variables. This~allows writing the differential of $\Lambda$ in the more compact form
\begin{equation}\label{ttttttt4}
d\Lambda =  -  \dfrac{1}{2} \sum_{x,y} \mathcal{K}^{xy} d(n_{xy}^2),
\end{equation}
where the coefficients $\mathcal{K}^{xy}$ form the symmetric $2\times 2$ matrix
\begin{equation}\label{astratto}
\mathcal{K}^{xy} =
\begin{bmatrix}
   \mathcal{B}^A & \mathcal{A}^{AB}   \\
    \mathcal{A}^{BA} & \mathcal{B}^{B}  \\
\end{bmatrix} \, .
\end{equation}

From \eqref{ttttttt4} it is immediate to prove \eqref{vbvbvb}, in fact one can easily see that 
\begin{equation}
\dfrac{\partial \Lambda}{\partial n_x^\nu} = \sum_y \mathcal{K}^{xy}n_{y\nu}, 
\end{equation}
leading to an explicit expression for the momenta in the matrix form
\begin{equation}
\begin{pmatrix}
\mu^A_\nu  \\
\mu^B_\nu \\
\end{pmatrix}
=
\begin{bmatrix}
   \mathcal{B}^A & \mathcal{A}^{AB}   \\
    \mathcal{A}^{BA} & \mathcal{B}^{B}  \\
\end{bmatrix} 
\begin{pmatrix}
n_{A\nu}  \\
n_{B\nu} \\
\end{pmatrix},
\end{equation}
which is equivalent to Equation~\eqref{vbvbvb}.

The distinction between the triangular and the square formulation is never explicitly discussed in the literature and the two are used interchangeably according to convenience. However it is important to keep the distinction clear in mind, because in the square formulation
\begin{equation}
\mathcal{A}^{AB} = - 2\dfrac{\partial \Lambda}{\partial n_{AB}^2},
\end{equation}
while in the triangle formulation
\begin{equation}
\mathcal{A}^{AB} = - \dfrac{\partial \Lambda}{\partial n_{AB}^2}.
\end{equation} 

There is no contradiction between the two, because in the first case the derivative is performed keeping $n_{BA}^2$ constant, while in the second case it is performed along the curve $n_{AB}^2=n_{BA}^2$, producing a double variation of $\Lambda$.

\section{The Relaxation-Time Approximation}\label{Rilassati}

Following~\cite{mihalas_book}, we assume that the photon distribution function $f$ is governed by the transport~equation
\begin{equation}\label{bolzi}
p^\mu \partial_\mu f = \sigma - \alpha f.
\end{equation}

We are working for simplicity in a flat space-time, with global inertial coordinates. We can interpret $\sigma$ as a source term, while $\alpha$ plays the role of an absorption coefficient (however it includes also a negative contribution coming from the stimulated emission, see e.g.,~\cite{clayton_book}). In this appendix, for simplicity, we will focus only on the thermal absorption/emission processes and we will ignore completely the scattering. It is clear that if the scattering is isotropic and coherent, it will just produce an additional term $\chi^e F^\nu$ to be included in the total force $G^\nu$.

If the photon gas was in local thermodynamic equilibrium (together with the matter element) its distribution function would be
\begin{equation}
f_{\text{eq}} = \dfrac{g}{h_p^3} \dfrac{1}{e^{-u_n^\mu p_\mu/\Theta^n}-1}.
\end{equation} 
where $g=2$ accounts for the spin degeneracy and $h_p$ is the Planck constant. Imposing the Kirchhoff law consists of assuming that 
\begin{equation}
\sigma = \alpha f_{\text{eq}}.
\end{equation}

The frequency $\nu$ of a photon measured in the matter rest-frame is related to the four-momentum $p^\mu$ through the relation
\begin{equation}
h_p \nu = -u_n^\mu p_\mu.
\end{equation}

This~relation can be used to prove that the specific intensity $I_\nu$ and the absorption opacity $\chi^A$ can be obtained using the relations~\citep{mihalas_book}
\begin{equation}
f= \dfrac{I_\nu}{h_p^4 \nu^3}  \spc \alpha= h_p\nu \chi^A.
\end{equation}

The coefficient $\chi^A$ in principle depends on the frequency, the~grey-body assumption consists of requiring that 
\begin{equation}
\chi^A(\nu)=\text{const}.
\end{equation} 

This~allows us to rewrite Equation~\eqref{bolzi} in the Anderson-Witting relaxation-time form~\citep{AndersonWitting1974,cercignani_book}
\begin{equation}\label{relax}
p^\mu \partial_\mu f =  - p^\mu   u_{n\mu} \dfrac{f_{\text{eq}} -f}{\tau_A},
\end{equation}
where
\begin{equation}
\tau_A = \dfrac{1}{\chi^A}
\end{equation}
is the relaxation time-scale towards equilibrium of the radiation gas.  

The generic moment of the radiation distribution function is defined as
\begin{equation}
\varphi^{\mu \alpha_1 ...\alpha_N} := \int p^{\alpha_1}...p^{\alpha_N}p^\mu f \dfrac{d_3 p}{p^0}.
\end{equation}

It is easy to show that \eqref{relax} implies
\begin{equation}
\partial_\mu \varphi^{\mu \alpha_1 ...\alpha_N} = -u_{n\mu} \dfrac{\varphi^{\mu \alpha_1 ...\alpha_N}_{\text{eq}} - \varphi^{\mu \alpha_1 ...\alpha_N}}{\tau_A}.
\end{equation}

Considering that
\begin{equation}
\varphi^\mu = \upgamma^\mu   \spc  \varphi^{\mu \alpha} = R^{\mu \alpha} ,
\end{equation}
we finally obtain
\begin{equation}
r_\upgamma = \chi^A (\hat{\upgamma}_{\text{eq}} -\hat{\upgamma}),
\end{equation}
where we have defined $\hat{\upgamma}=-u_{n\mu}\upgamma^\mu$, and
\begin{equation}
G^\mu = \chi^A (\hat{\varepsilon}-\hat{\varepsilon}_{\text{eq}})u_n^\mu + \chi^A F^\mu.
\end{equation}

\section{Calculations}

In this appendix we report in detail the calculation which were omitted from the main text.

\subsection{Onsager Symmetry of the Dissipative Coefficients}\label{appendixB}

We consider a homogeneous matter-radiation multifluid prepared in an initial state such that both the components are at rest. Under this condition we can combine Equations \eqref{dRho} and \eqref{dVarepsilon} to write the differential of the total entropy as
\begin{equation}
ds = \dfrac{d\rho}{\Theta^n} -\dfrac{\mu}{\Theta^n} dn + \dfrac{d\varepsilon}{\Theta^\upgamma} + \dfrac{\mathbb{A}^\upgamma}{\Theta^\upgamma} d\upgamma.
\end{equation}

Imposing the conservation of energy and particle number,
\begin{equation}
dn=0 \spc d\varepsilon=-d\rho,
\end{equation}
we obtain the differential
\begin{equation}
ds= \bigg( \dfrac{1}{\Theta^n}-\dfrac{1}{\Theta^\upgamma} \bigg)d\rho +\dfrac{\mathbb{A}^\upgamma}{\Theta^\upgamma} d\upgamma.
\end{equation}

Therefore we are able to identify the thermally fluctuating variables
\begin{equation}
y_1 = \rho  \spc y_2=\upgamma
\end{equation} 
and their conjugates
\begin{equation}
C^1=\dfrac{1}{\Theta^n} -\dfrac{1}{\Theta^\upgamma}  \spc C^2 = \dfrac{\mathbb{A}^\upgamma}{\Theta^\upgamma}.
\end{equation}

Onsager principle states that if we write the evolution of the variables $y_A$ in the form
\begin{equation}\label{dcfvbgnhjm}
\dot{y}_A = \sum_B L_{AB}C^B,
\end{equation}
then $L_{AB}$ is symmetric.

From \eqref{rgaMMa} and \eqref{calormio} we find that in homogeneous configurations and in the absence of a relative~flow
\begin{equation}
\dot{\rho}= Q  \spc \dot{\upgamma} = r_\upgamma,
\end{equation}
which recalling \eqref{EEEEEspandiamo}, gives the formulas
\begin{equation}
\begin{split}
& \dot{\rho} = k_\upgamma \mathbb{A}^\upgamma+ k_T (\Theta^\upgamma -\Theta^n) \\
& \dot{\upgamma} = \Xi_{\upgamma \upgamma}\mathbb{A}^\upgamma + \Xi_{\upgamma T} (\Theta^\upgamma -\Theta^n).  \\
\end{split}
\end{equation}

It is easy to rewrite this system in the form \eqref{dcfvbgnhjm}:
\begin{equation}
\begin{split}
& \dot{y}_1 =\Theta^\upgamma k_\upgamma C^2 +\Theta^\upgamma \Theta^n k_T C^1 \\
& \dot{y}_2 =\Theta^\upgamma \Xi_{\upgamma \upgamma}C^2 +\Theta^\upgamma \Theta^n \Xi_{\upgamma T} C^1, \\
\end{split}
\end{equation}
which produces the reciprocal relation
\begin{equation}
k_\upgamma = \Theta^n \Xi_{\upgamma T}.
\end{equation}

\subsection{Computation of the Bulk Viscosity Coefficient}\label{qazsedc}

The chemical-transfusion matrix is
\begin{equation}
\Xi_{AB}
=
\begin{bmatrix}
   \Xi_{\upgamma \upgamma} & \Xi_{\upgamma T}  \\
    k_\upgamma/\Theta^n & k_T/\Theta^n  \\
\end{bmatrix}.
\end{equation}

The inverse chemical matrix is, therefore,
\begin{equation}\label{gasp!}
\Xi^{AB} = \dfrac{\Theta^n}{\Xi_{\upgamma \upgamma}k_T - \Theta^n\Xi_{\upgamma T}^2}
\begin{bmatrix}
   k_T/\Theta^n & -\Xi_{\upgamma T}   \\
   -\Xi_{\upgamma T} & \Xi_{\upgamma \upgamma}  \\
\end{bmatrix}. 
\end{equation}

The equilibrium fractions are given by
\begin{equation}
x_\upgamma^{\text{eq}} =\dfrac{\upgamma}{n}\bigg|_{\mathbb{A}^\upgamma =0, \Theta^n=\Theta^\upgamma} \spc  x_{sn}^{\text{eq}} =\dfrac{s_n}{n}\bigg|_{\mathbb{A}^\upgamma =0, \Theta^n=\Theta^\upgamma} .
\end{equation}

We note that the relation \eqref{contrainatoygviyg} implies
\begin{equation}\label{gasp22}
\dfrac{\partial x^{\text{eq}}_{sn}}{\partial v}\bigg|_{x_s} = -b_R \dfrac{\partial x^{\text{eq}}_\upgamma}{\partial v} \bigg|_{x_s}.
\end{equation}

Plugging \eqref{gasp!} and \eqref{gasp22} into \eqref{ZZZeTa}, we obtain
\begin{equation}
\zeta= \dfrac{k_T +2b_R\Theta^n \Xi_{\upgamma T}+b_R^2 \Theta^n \Xi_{\upgamma \upgamma} }{\Xi_{\upgamma \upgamma}k_T - \Theta^n\Xi_{\upgamma T}^2} \bigg( \dfrac{\partial x^{\text{eq}}_\upgamma}{\partial v} \bigg|_{x_s} \bigg)^2.
\end{equation}

\subsection{Eckart Decomposition of the Entropy Current}\label{perchenolinonso}

From the definition \eqref{Ecktemp} we have
\begin{equation}
s_E = s_n +s_\upgamma \Gamma_{n\upgamma}  \spc  \Theta_E = \Theta^\upgamma \Gamma_{n\upgamma},
\end{equation}
which plugged into \eqref{entrcurr}, gives
\begin{equation}
s^\nu = (s_n +s_\upgamma \Gamma_{n\upgamma})u_n^\nu + \dfrac{q^\nu}{\Theta^\upgamma \Gamma_{n\upgamma}}.
\end{equation}

Thus, in the tetrad which comoves with the matter element, we have
\begin{equation}\label{1234567}
q^j = \Theta^\upgamma \Gamma_{n\upgamma} s^j. 
\end{equation}

On the other hand, in this basis $s_n^j=0$, so
\begin{equation}\label{9876543}
s^j = s_\upgamma \Gamma_{n\upgamma} v^j.
\end{equation}

The equation of state \eqref{frgthy} and the Legendre transform \eqref{PRRraDD} imply that in chemical equilibrium ($\mathbb{A}^\upgamma=0$) we have the well-known formula
\begin{equation}
s_\upgamma \Theta^\upgamma = \dfrac{4}{3} \varepsilon,
\end{equation}
which combined with \eqref{1234567} and \eqref{9876543}, gives
\begin{equation}
q^j = \dfrac{4}{3} \varepsilon \Gamma_{n\upgamma}^2  v^j.
\end{equation}

By comparison with \eqref{zoooooom}we obtain
\begin{equation}
q^\nu =F^\nu.
\end{equation}

\section{Radiation-Mediated Bulk Viscosity of Non-Degenerate Ideal Gases}\label{loabbiamofatto}

Let us consider Equation~\eqref{fvfbgnhm} and let us assume that the matter fluid is a non-degenerate ideal gas. Since all the coefficients in the formula for $\zeta$ are computed in equilibrium, in this appendix we will impose $\Theta^n=\Theta^\upgamma=\Theta$ and $s_\upgamma = b_R \upgamma$. Therefore we can write
\begin{equation}\label{rimpluzzo}
    x_{sn} = \log(v\Theta^w) + \text{const} \spc x_{s\upgamma} = \dfrac{4}{3} a_R v \Theta^3,
\end{equation}
where $w=3/2$ if the matter-fluid is a non-relativistic gas and $w=3$ if it is an ultra-relativistic gas. Now, we immediately see that in the latter case
\begin{equation}
x_s =x_s(v\Theta^3),
\end{equation}
thus the adiabatic curves are given by
\begin{equation}
    v\Theta^3 =\text{const}.
\end{equation}

Considering that
\begin{equation}
x_\upgamma = \dfrac{4 a_R}{3 b_R} v\Theta^3,
\end{equation}
we find
\begin{equation}
    \dfrac{\partial x^{\text{eq}}_\upgamma}{\partial v} \bigg|_{x_s} =0
\end{equation}
and, therefore,
\begin{equation}
    \zeta =0.
\end{equation}

This~result is in agreement with the well-known fact that in ultra-relativistic ideal gases the bulk viscosity is identically zero \citep{Weinberg1971,BulkGavassino}.

Let us focus on the case $w=3/2$ (non-relativistic matter-gas). Starting from the obvious relation
\begin{equation}
 \dfrac{\partial x_s}{\partial v} \bigg|_{x_s} =  \dfrac{\partial x_{sn}}{\partial v} \bigg|_{x_s} + \dfrac{\partial x_{s\upgamma}}{\partial v} \bigg|_{x_s} = 0,
\end{equation}
we obtain, recalling \eqref{rimpluzzo},
\begin{equation}\label{bazZinga}
  \dfrac{\partial \Theta}{\partial v} \bigg|_{x_s}= -\dfrac{2\Theta}{3 v} \, \dfrac{1+x_{s\upgamma}}{1+2x_{s\upgamma}}. 
\end{equation}

This~can be used to show that
\begin{equation}
    \dfrac{\partial x^{\text{eq}}_\upgamma}{\partial v} \bigg|_{x_s} = - \dfrac{\upgamma}{1+2x_{s\upgamma}},
\end{equation}
which plugged into \eqref{fvfbgnhm}, gives
\begin{equation}
    \zeta = \dfrac{4 P_\upgamma}{3\chi^A (1+2x_{s\upgamma})^2}.
\end{equation}

It is useful to rewrite this formula using more standard notation. To do this, we introduce the pressure ratio
\begin{equation}
    \alpha_P = \dfrac{P_\upgamma}{P_n} = \dfrac{x_{s\upgamma}}{4},
\end{equation}
where the second identity follows from the ideal gas assumption $P_n = n \Theta$. Plugging it into \eqref{bazZinga} we recover the well-known formula \citep{mihalas_book}
\begin{equation}
    \dfrac{\partial \log \Theta}{\partial \log v} \bigg|_{x_s} = - \dfrac{1+4 \alpha_P}{3/2 + 12 \alpha_P}.
\end{equation}

Finally, our expression for the bulk viscosity becomes
\begin{equation}
    \zeta = \dfrac{4P_n}{3\chi^A}  \, \dfrac{\alpha_P}{(1+8\alpha_P)^2}.
\end{equation}

We see that the second fraction tends to suppress $\zeta$ as  $\alpha_P \longrightarrow 0$ or $\alpha_P \longrightarrow +\infty$. Intuitively, this is due to the fact that since the bulk viscosity is due to the dissipative processes which tend to equilibrate the temperatures of matter and radiation (which would depart from each other during a fast expansion), it becomes important only if both the species give a relevant contribution to the overall stress-energy tensor.

\bibliographystyle{mnras}
\bibliography{Biblio}

\end{document}